# Pathologies of Quenched Lattice QCD at non–zero Density and its Effective Potential

Maria–Paola Lombardo

*HLRZ c/o KFA, D-52425 Jülich and DESY, D-22603 Hamburg, Germany*

John B. Kogut

*Physics Department, University of Illinois at Urbana-Champaign, Urbana, IL 61801-30*

D. K. Sinclair

*HEP Division, Argonne National Laboratory, 9700 South Cass Avenue, Argonne, IL 60439*

(November 23, 1995)

## Abstract

We simulate lattice QCD at non–zero baryon density and zero temperature in the quenched approximation, both in the scaling region and in the infinite coupling limit. We investigate the nature of the forbidden region – the range of chemical potential where the simulations grow prohibitively expensive, and the results, when available, are puzzling if not unphysical. At weak coupling we have explored the sensitivity of these pathologies to the lattice size, and found that using a large lattice ($64 \times 16^3$) does not remove them. The effective potential sheds considerable light on the problems in the simulations, and gives a clear interpretation of the forbidden region. The strong coupling simulations were particularly illuminating on this point.

Typeset using REVTeX

# I. INTRODUCTION

The absence of a simulation algorithm for finite density QCD is an outstanding problem both of QCD thermodynamics and lattice field theory [1]. From the lattice viewpoint, the source of these difficulties is the complex nature of the QCD finite density Action [2], [3]. This property prevents the use of naive probabilistic methods in evaluating the functional integral, thus calling either for exact studies, which are extremely compute intensive, or for suitable approximations.

Many studies, especially the early ones [2], [4], have pursued approximations. Finite density QCD has been studied in the quenched approximation, which is the most widely exploited approximation in lattice QCD: this approximation has problems in the chiral limit because it ignores the axial anomaly, but it is satisfactory for many practical calculations. It was therefore rather perplexing and disappointing when severe problems were first reported in quenched simulations of finite density QCD [4]: at finite quark mass, the onset of chiral symmetry restoration appeared to occur at a chemical potential of half the pion mass, which would extrapolate to zero in the chiral limit. Several explanations of this seemingly unphysical result were proposed, but none of them led to a satisfactory treatment of finite density QCD, nor did they indicate if the quenched approximation was at fault (except in that they relied on simple probabilistic arguments that failed for the complex action of full QCD at finite chemical potential) or if there were intrinsic problems in the lattice formulation of fermions and chemical potential that would survive more sophisticated treatments.

In Ref. [5], we attempted to confirm the relationship $\mu_{onset} = m_\pi/2$ in the quenched model. We considered that to be necessary because, as suggested by other authors in the past, the coincidence of the onset with one half the pion mass might only have been a numerical accident, the correct relationship being $\mu_{onset} = m_N/3 - \Delta$, where $\Delta$ is the contribution of the nuclear binding energy. If this was the case, the problems with finite density would not have been serious. These considerations called for high precision measurements, never performed before. Later arguments [6] indicating that the transition should indeed occur precisely at $\mu = m_\pi/2$ lack rigor and could well do with the scrutiny of such measurements. This work, alas, confirmed the onset at $\mu = m_\pi/2$.

However, as discussed at length in Ref. [5], the finite temporal extent of the lattice might be responsible for unphysical closed quark loops winding around the system. Such loops can generate huge fluctuations and give misleading indications of deconfinement and chiral symmetry restoration. These considerations motivated us to follow-up Ref. [5] with simulations on a lattice of greater temporal extent. We discuss these results in Section II below. In particular, we shall give an overview of the results in Section II.A, while Sections II.B and II.C , are devoted to the spectrum analysis and finite size effects, and can be skipped by the reader uninterested in technical details. We next turn to the strong coupling limit of the theory. Section III is devoted to new simulations performed in that regime. The strong coupling limit has little phenomenological relevance, but it displays both confinement and chiral symmetry breaking – the QCD characteristics which are relevant in this context. There are several considerations which made this study appealing to us: first, in the same spirit of the weak coupling simulations, we wanted to have a clean measure of the onset: at strong coupling half the pion mass and one third the baryon mass are far apart, and are easily distinguished numerically; second, the simulations are much less computer intensive, and accurate results can be obtained; finally, several analytic treatments are available, which can clarify the interpretation of the results. We have also introduced new observables which have helped clarify some pathologies, and have helped us arrive at a coherent description of the failure of the quenched approximation. We discuss these measurements in Section III.A. Section III.B contains the ordinary thermodynamics results. They were particularly clear, and were easily extended to $\mu > m_N/3$, the saturation region. Finally, in Section III.C we contrast the results of the simulations with the analytic predictions. This gives a new interpretation of the forbidden region, $\mu > m_\pi/2$.

## II. QCD AT NON–ZERO BARYON DENSITY IN THE SCALING REGION: IMPROVED OPERATORS, TEMPERATURE EFFECTS AND WINDING LOOPS.

This section contains an upgrade of our previous work on quenched QCD at non-zero baryon-number density and $\beta = 6.0$ [5]. We briefly summarize here our main findings and open problems, and refer to [5] for introductory material and a general review. Our best success there was a measurement of the nucleon mass for $\mu < m_\pi/2$. The results were in agreement with the predictions of a simple constituent quark model, and hinted at a critical value for the chemical potential given by one third the baryon mass. Unfortunately, the more interesting region $\mu > m_\pi/2$ evaded us: enormous fluctuations dominated the results, and not only was it impossible to estimate the mass spectrum, but the shape of the propagators themselves were grossly distorted. The results for thermodynamic quantities were somewhat better, and we were able to detect a modest decrease in the chiral condensate at $\mu \simeq m_\pi/2$.

The main purpose of this new work is thus twofold: first, to understand and possibly eliminate the source of the anomalous fluctuations in the spectrum, in order to extend our simulations inside the interesting region of the chemical potential; second, to explore the possibility of residual temperature (the temperature being the reciprocal of the time extent) effects on the thermodynamic observables. Both of the above called for a simulation on a lattice of larger temporal extent: this pushes the temperature further toward zero, and possibly suppresses precocious, unphysical quark loops winding around the lattice in the time direction. We have also proposed another possible explanation for the pathologies afflicting the spectrum: the wall source used for the spectrum measurements could produce strong distortions in the propagators. Remember that such a source, used in conjunction with with a point sink, alters the value of the coefficients $a_i$ in the expansion of a propagator $G(t)$, $G(t) = \sum a_i e^{m_i t}$, by maximizing $a_0$, the projection onto the fundamental state, a clear numerical advantage. Unfortunately, this advantage is accompanied by a possible lack of positivity of the other coefficients, leading to instabilities in the results, amplified by the asymmetries introduced by the chemical potential. In the new set of simulations we have thus decided to use a noisy–wall source: namely, a random $SU(3)$ matrix was put on the (even,even,even) sites of the spatial cube $t = 0$, and the Dirac operator was inverted in this background. It is easy to show that the average local-hadron propagators are the same as the ones obtained with a point source, thus avoiding the positivity problems we have just mentioned. There is a statistical gain due to the increase of the number of source points (roughly, its square root), and a loss because of possible incomplete cancelations of gauge-variant terms which are inherent in this technique [9]. The real advantages can only be judged a posteriori. A sure disadvantage, well known from the zero chemical potential simulations, is the absence of useful results for the baryon – with this source, asymptotic behavior sets in only at a large time interval where the baryon propagator is too small to measure. (N.B. we considered using a version of the noisy–wall source which corresponds to smeared sources, along with the corresponding smeared sinks. However, this gives less statistical gain than for the point version.)

Summarizing, we have doubled the temporal size of the lattice ($16^3 \times 64$ vs. $16^3 \times 32$), and used a noisy–wall (as opposed to a rigid wall) source for the spectrum measurements. Making these two changes (source, and time extension) simultaneously was the best way to optimize our computer resources, but unfortunately we were not able to disentangle the main sources of the troubles afflicting previous spectrum computations. The measurement strategy for the order parameter and number density was the same as before, so in these cases we are genuinely monitoring finite temperature effects.

As in past work, we exploited the "global" $Z_3$ symmetry of quenched QCD, using three different anti-periodic boundary conditions for the fermions in the time direction: $\psi_i(N_t) = -Z_i\psi(0)$, where $Z_i$ stands for each of the three cube roots of unity. Remember that this can help enforce the constraints of confinement by suppressing unphysical winding loops. It is thus interesting to monitor the statistical gain coming from averaging over these boundary conditions as a function of the lattice time extent. The bigger the gain, the more important the role of the (unphysical) winding loops, so the interplay of temperature and winding loops can be clearly exposed.

We have generated 30 independent configurations on a $16^3 \times 64$ lattice from a code which is a blend of Metropolis and over-relaxed algorithms. A subset of these configurations was used for the inversion of the Dirac operator for each $\mu$ value. The quark mass was .02, as in our previous work. We have inverted the Dirac operator for $\mu = 0.0, .1, .15, .17, .2$ on $13, 13, 26, 12, 28$ configurations, respectively. We have also done a few exploratory runs for larger $\mu$ values ($\mu \simeq 1.$), searching for hints of saturation (see [4], and the following Section), but the number of iterations required for the inversion was always huge, and the number density rather low. In the free theory saturation occurs only for rather large $\mu$ : we probably observed a similar trend at $\beta = 6.0$, in agreement with the observations of [16].

The measurements and analyses were exactly the same as in our past work, to which we refer the reader for details.

### A. Overview of the results

The results for the chiral condensate and the number density are shown in Table I and II and collected in Figs. 1 and 2. They are systematically compared with those obtained on the smaller lattice (for the sake of comparison, we have repeated the analysis on the smaller lattice including all the data, without any cuts, which accounts for some difference (statistically irrelevant) among the results in this table and the ones of Ref. [17]).

For small values of the chemical potential ($\mu < m_\pi/2$) we do not see any temperature effect for the chiral condensate, while for $\mu > m_\pi/2$ the overall behavior does not exclude the possibility of (modest) finite temperature effects. We will come back to this point in the last subsection.

In the spectrum the most visible phenomenon concerns the pion propagator. The huge and unphysical fluctuations described above in our past measurements [17] disappear. However, the number of iterations required for the inversion when $\mu$ exceeds $m_\pi/2$ remains big, of order $10^4$, meaning that the Dirac operator is still nearly singular.

| | $<\bar{\psi}\psi>$ | |
|---|---|---|
| $\mu$ | $16^3 \times 32$ | $16^3 \times 64$ |
| 0.0 | .1377(6) | .1376(2) |
| .1 | .1375(8) | .1376(3) |
| .15 | .1362(18) | .1381(5) |
| .17 | .1359(140) | .1357(19) |
| .20 | .1088(100) | .1277(35) |

TABLE I. Results for the chiral condensate as a function of the chemical potential on the $16^3 \times 32$ and on the $16^3 \times 64$ lattices. We quote the average over the three boundaries and the two opposite $\mu$ values

| | $<J_0>$ | |
|---|---|---|
| $\mu$ | $16^3 \times 32$ | $16^3 \times 64$ |
| 0.0 | .00016(93) | .00029(50) |
| .1 | .00033(83) | .00011(56) |
| .15 | .00069(145) | -.00013(52) |
| .17 | .00071(161) | .00028(212) |
| .20 | .00510(899) | .00481(315) |

TABLE II. Results for the number density as a function of the chemical potential on the $16^3 \times 32$ and on the $16^3 \times 64$ lattices. We quote the average over the three boundaries and, for $\mu \neq 0$, over the two opposite $\mu$ values

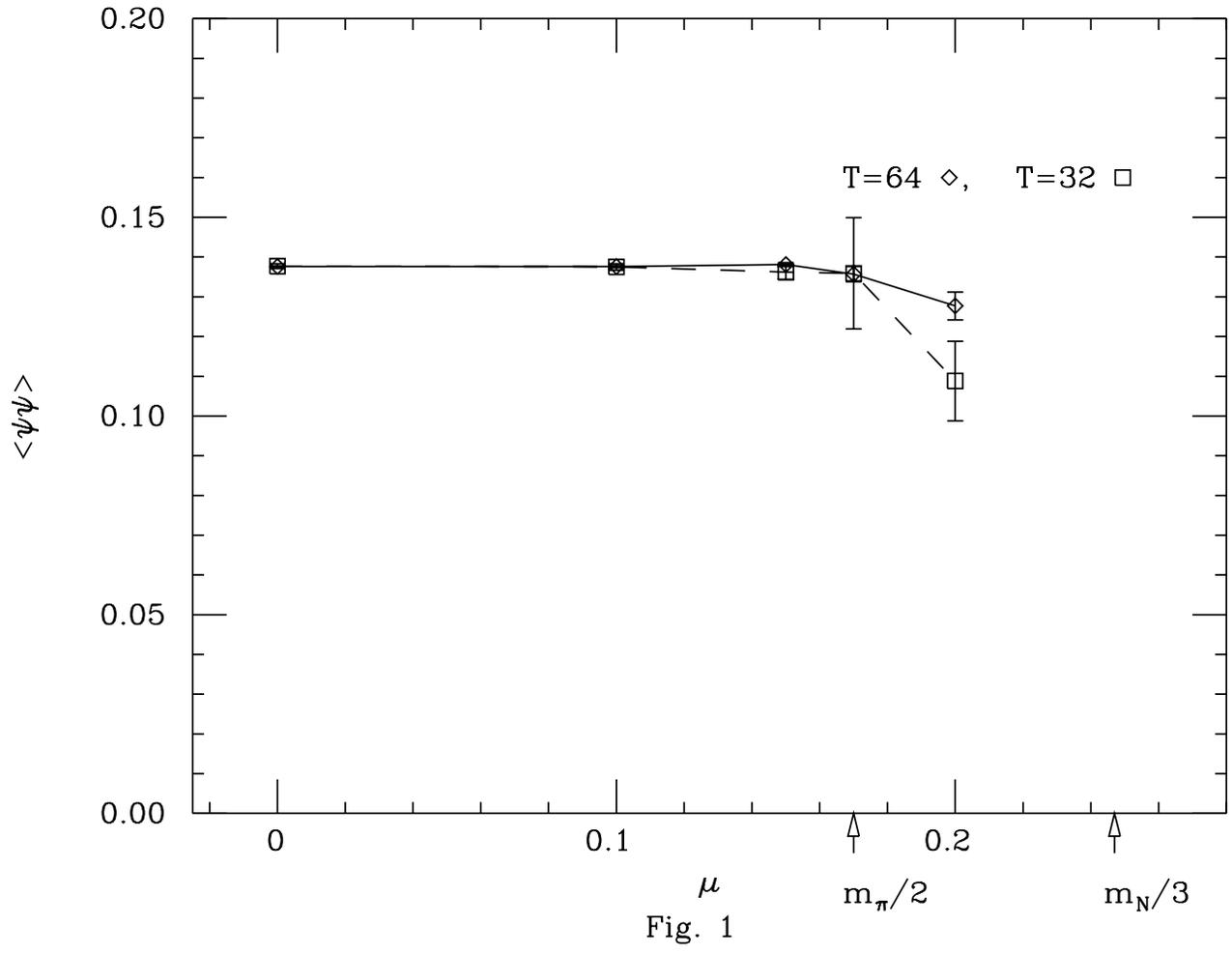

Figure 1. Results for the chiral condensate at $\beta = 6.0$ (diamonds). The results on the $16^3 \times 32$ lattice are shown for comparison (squares)

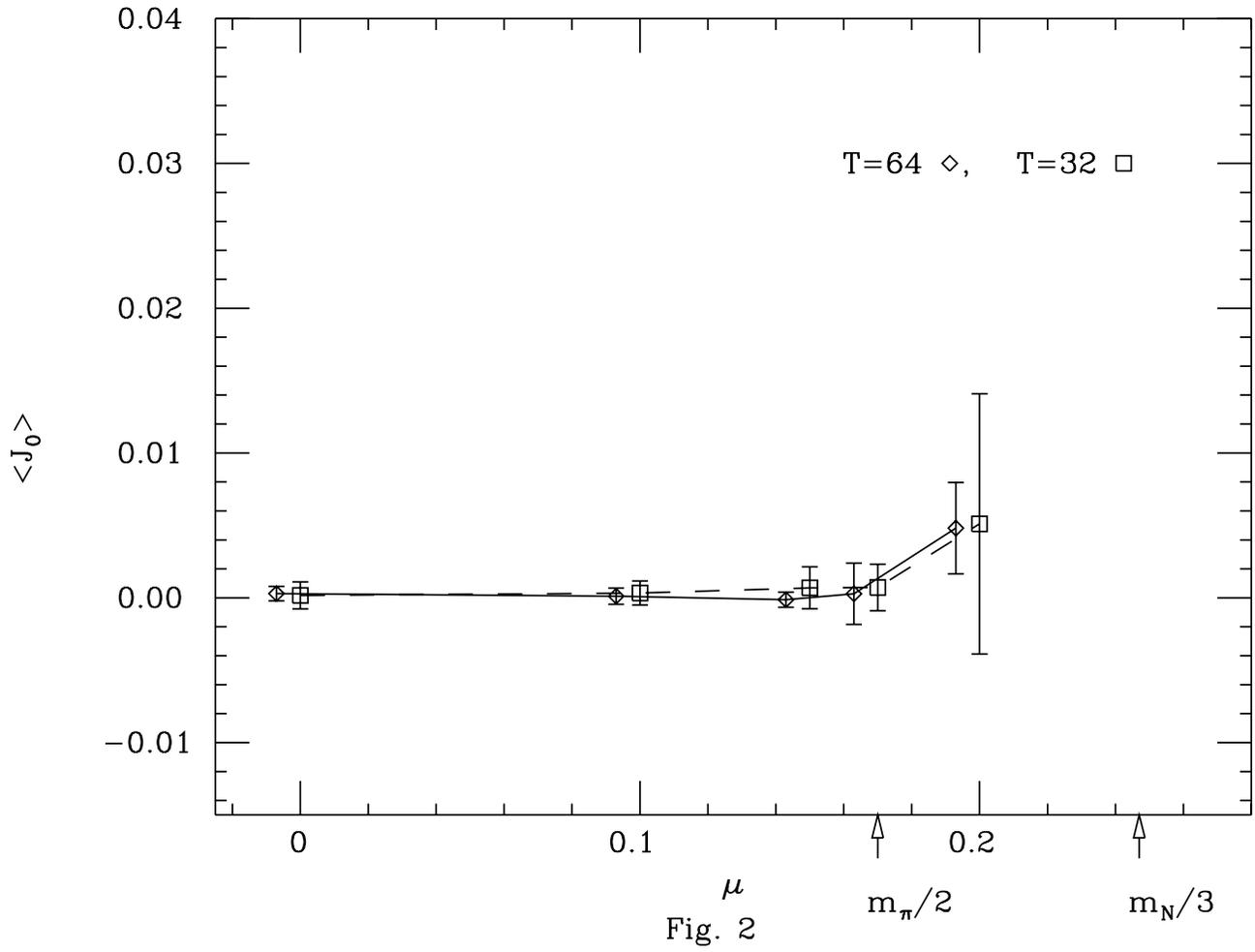

Figure 2. Results for the number density at $\beta = 6.0$ (diamonds). The results on the $16^3 \times 32$ are shown as squares, slightly displaced.

If we ascribe the improvement to the different sources we were using, we would conclude that the source affects the amplitudes, but not the poles, as expected. Despite the better behavior of the propagators, the pion mass estimate was not compelling for $\mu > m_\pi/2$, while for smaller chemical potentials we confirm previous results. We shall give some detail on the mass analysis in the next subsection. Here, we quote the results for the pion mass in Table III, again comparing with the results on the smaller lattice. We see that there are no discernible temperature effects. Also, remember that we were using two different sources, wall, and point with the noisy improvement. For our wall source, the effective mass estimator appears to reach its asymptotic behavior from below (see the details in the analysis Section below). In particular, as we will show in more detail later, our mass estimate at $\mu = .17$ is certainly non–asymptotic and should be considered as a lower bound. On the other hand, because of the positivity properties we mentioned above, the effective mass estimates from the propagators obtained with noisy–wall source give upper bounds. One might be tempted to "average" the results from the two different sources, thus obtaining a nearly constant pion mass across the transition.

We summarize all our findings for the spectrum in Fig. 3. Besides the old results, and the new ones from Table III, we show there the estimates for the pion mass at $\mu = .2$ (see next section for discussions and caveats), the point with the huge error coming from the full sample, the other one, slightly displaced, from a subset from which the most wildly fluctuating propagators have been excluded (in the preliminary results reported in [17] the statistical sample was smaller, so the cut was mandatory and we did not have an estimate from the full configuration ensemble). The fairest statement is that, despite our attempts, we do not have a reasonable mass estimate at $\mu = .2$, even though the propagator was better behaved than the one on the smaller lattice.

It would be interesting to determine if the pion mass does indeed increase for $\mu > m_\pi/2$, which would be an indication of chiral symmetry restoration. Alternatively, we can search for hints of chiral symmetry restoration in the spectrum sector by looking at the propagators of the chiral partners. We have indeed observed (Fig. 4) that the scalar and pseudoscalar propagators tend toward degeneracy. The same holds in the vector–pseudovector channel, but in both cases the effect is not dramatic, given that the two propagators are very close even at $\mu = 0$. Moreover, because of the different sources we were using on the smaller lattice, a finite size study of these quantities is not possible at this time.

| $\mu$ | $m_\pi$ | |
|---|---|---|
| | $16^3 \times 32$ | $16^3 \times 64$ |
| 0.0 | .3396(36) | .3379(31) |
| .1 | .3374(75) | .3440(51) |
| .15 | .3182(52) | .3464(142) |
| .17 | .313(15) | .41(3)(4) |

TABLE III. Pion masses as a function of $\mu$ on the two lattices. See text for details

| | $m_\pi$ | | | | | |
|---|---|---|---|---|---|---|
| | 0.0 | .1 | .15 | .17 | .20 | .20 (cut) |
| F2 [5–32] | .3379(31) | .3440(51) | .3464(142) | – | – | – |
| Em [20–30] | .3356(42) | .3376(47) | – | – | – | – |
| F2 [2–9] | .3610(55) | .3510(55) | .3609(57) | .4107(324) | .58(29) | .32(13) |
| F2 [3–7] | .3913(88) | .3914(87) | .3807(104) | .4155(433) | .60(24) | .37(24) |
| Em [14–16] | .3440(45) | .3442(52) | .3307(584) | – | – | – |

TABLE IV. Estimates for the pion mass from effective masses (Em) or two particle fits (F2). The first two lines are the "good" results

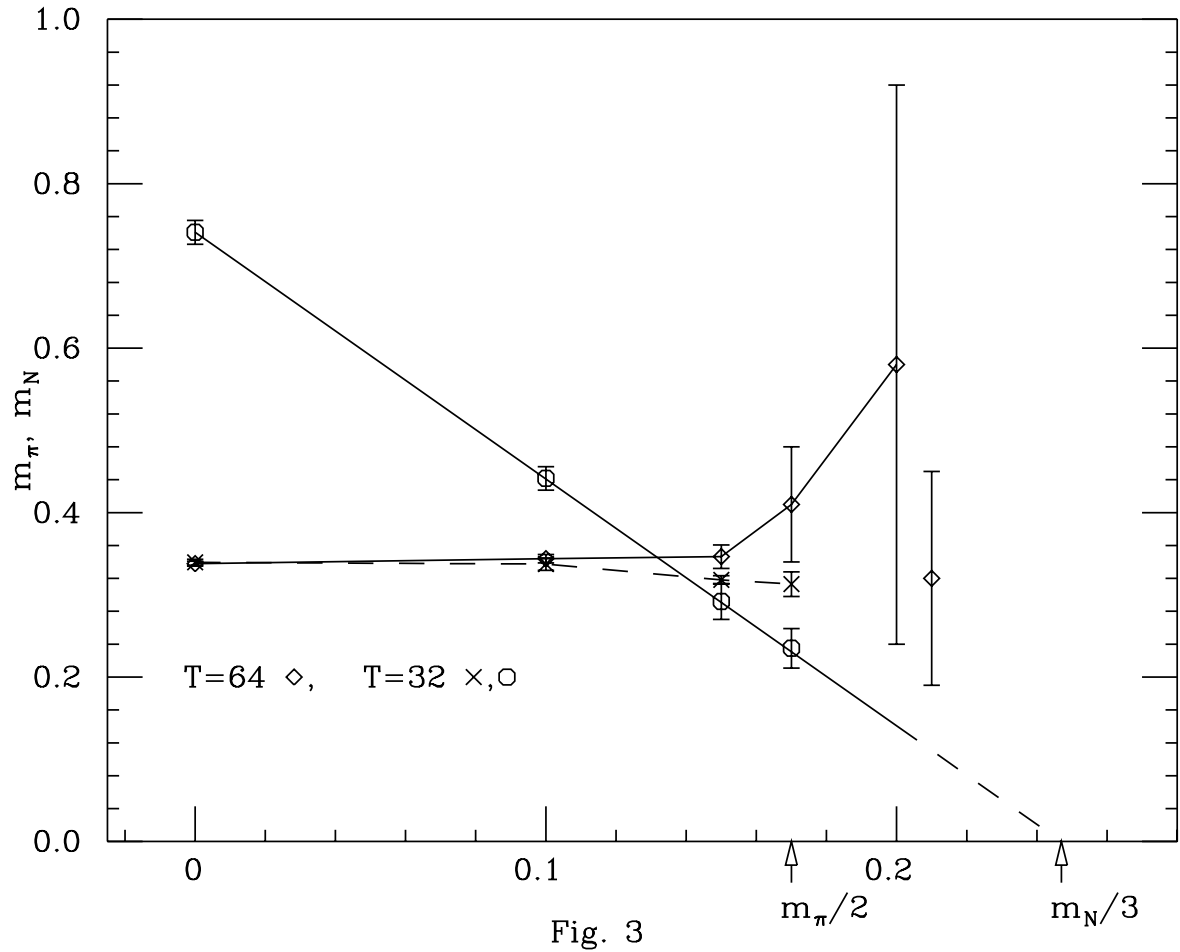

Figure 3. Summary of the spectroscopic results on the two lattices. See text for details.

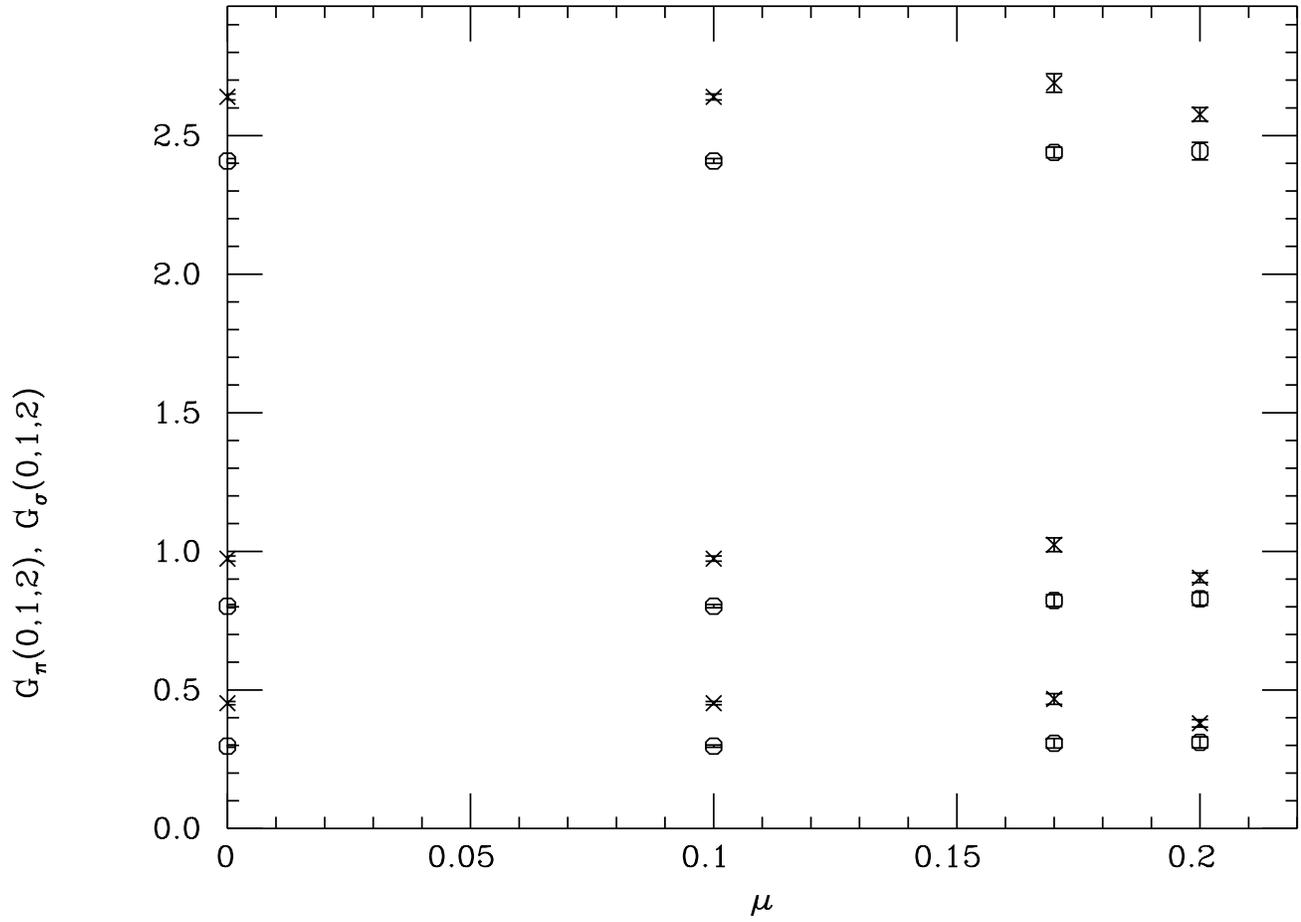

Figure 4. Scalar (circles) and pseudoscalar (crosses) propagators as a function of $\mu$. The (upper, middle, lower) couple of values is for the propagators at time separation (0,1,2).

In conclusion, we have observed only very modest hints of chiral symmetry restoration in the spectrum and in the thermodynamics, possibly affected by residual temperature effects. We have shown how suitable observables can help control major pathologies. However, these results alone do not determine the physical nature of the onset at $\mu = m_\pi/2$.

### B. Details of the spectrum analysis

We present here some details of our spectrum analysis. They provide some information on the qualitative effects of the chemical potential on the propagators.

We concentrate on the pion propagator, which is the best-behaved. The baryon propagator loses its numerical significance at distance 1–2 and we cannot extract useful information from it.

We show in Fig. 5 the collection of the effective mass plots for the pion (later in this work we will also extract masses from the t-dependence of the propagators). The quality of the data is good for small $\mu$, and plots, such as Fig. 6 for $\mu = .1$, show good agreement between our mass estimates obtained from the two lattice sizes, and the two source methods. As $\mu$ is increased, the signal is lost earlier and earlier in time. A better idea of the quality of the results can be obtained from Table IV, where we give the mass estimates from different analyses. It is clear that the asymptotic behavior for the effective mass sets is only for $t > 20$ (considering the behavior of $\mu = .0$ and $\mu = .1$), so the effective mass analysis does not give results for $\mu > .1$.

Next we considered two particle fits (fundamental + excited). In this case we were able to use time separations as small as five, although large time distances were still needed to stabilize the results. Fig. 7 demonstrates the quality of the fits, which were successful up to $\mu = .15$.

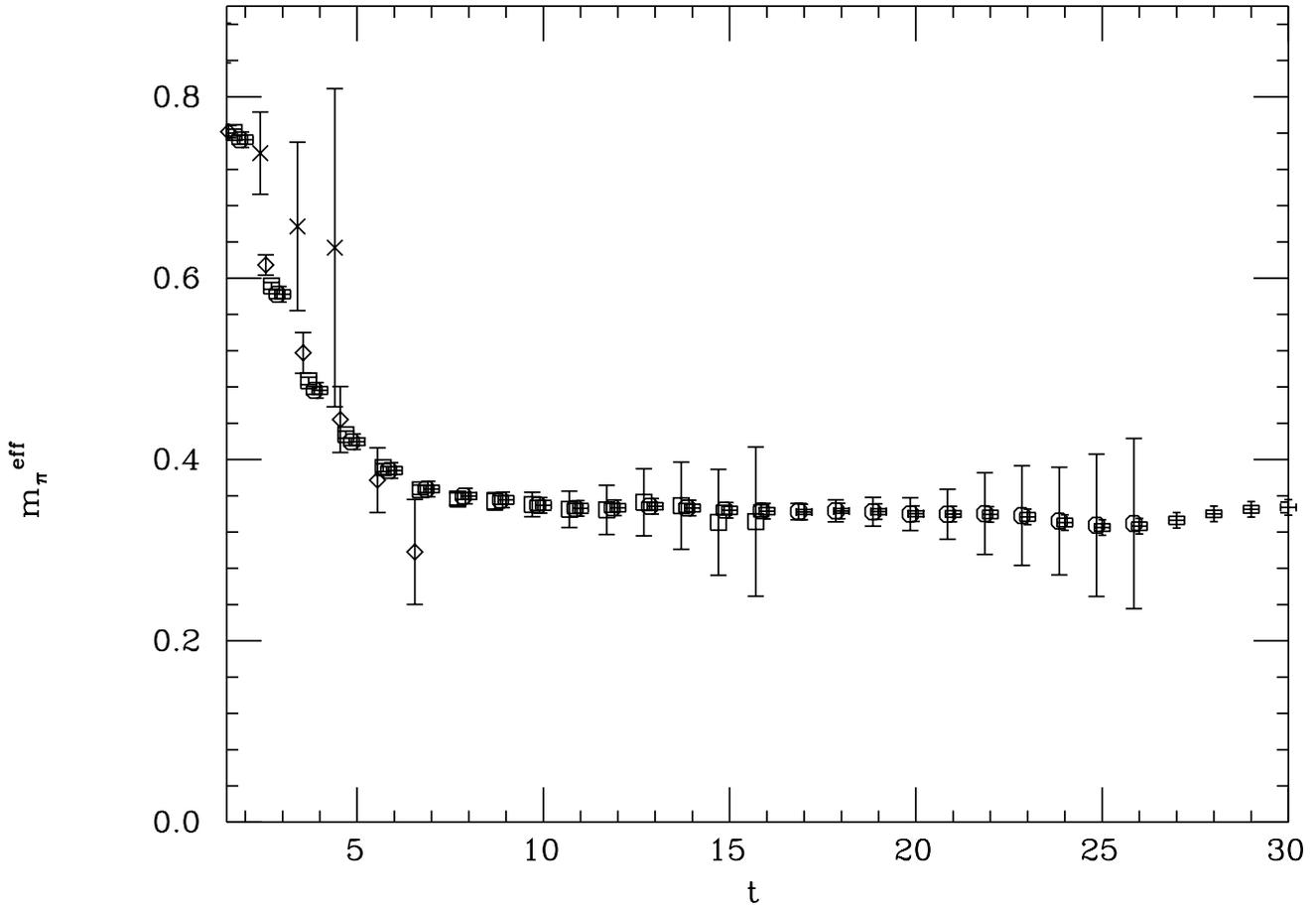

Figure 5. Effective masses as a function of time for $\mu = (0.0, .1, .15, .17, .2)$ (pluses, circles, squares, diamonds, crosses). Data for different $\mu$'s are slightly displaced.

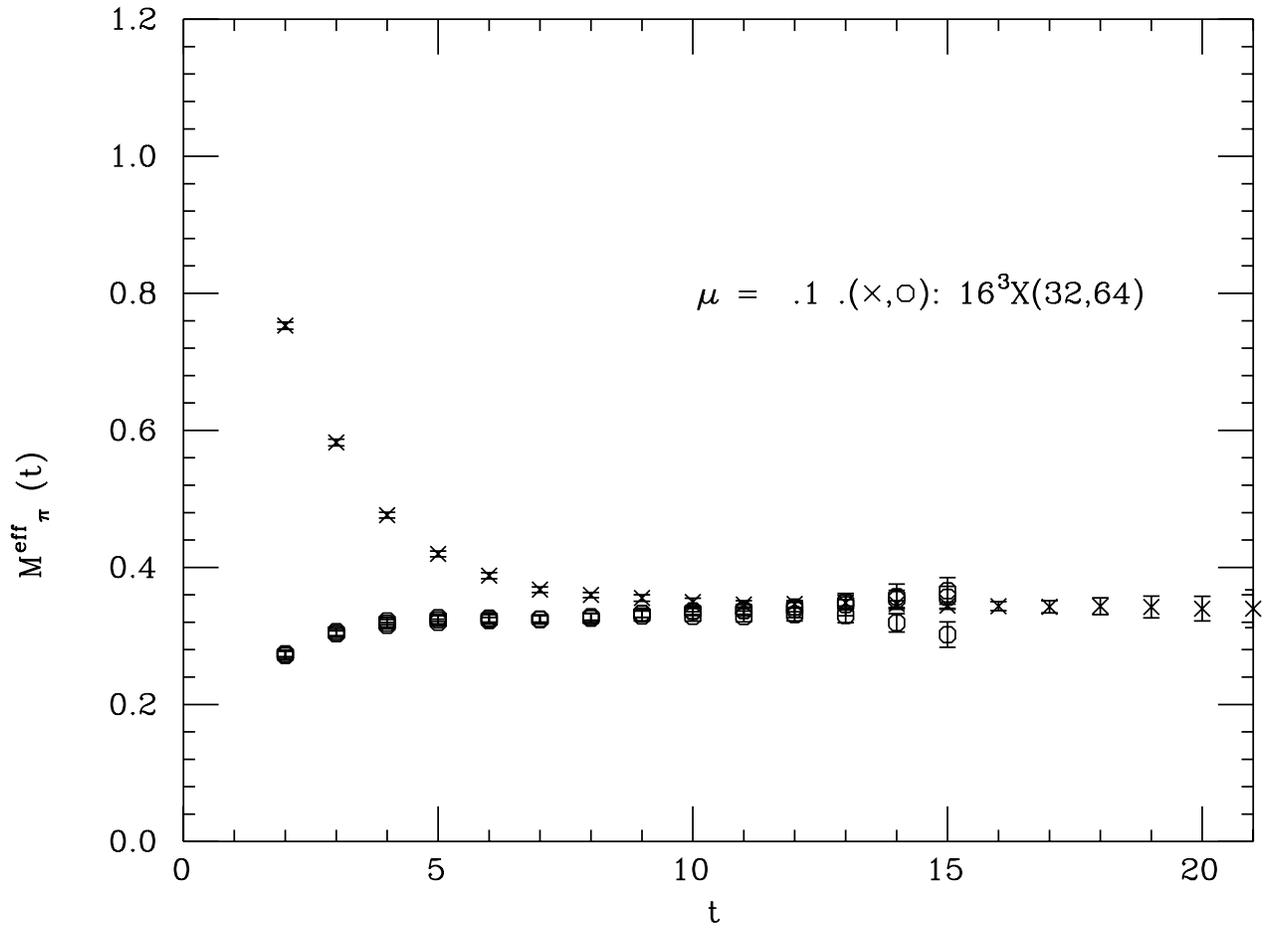

Figure 6. The results for the effective mass at $\mu = .1$ on the big lattice are contrasted with the ones on the smaller lattices, for the three different boundaries.

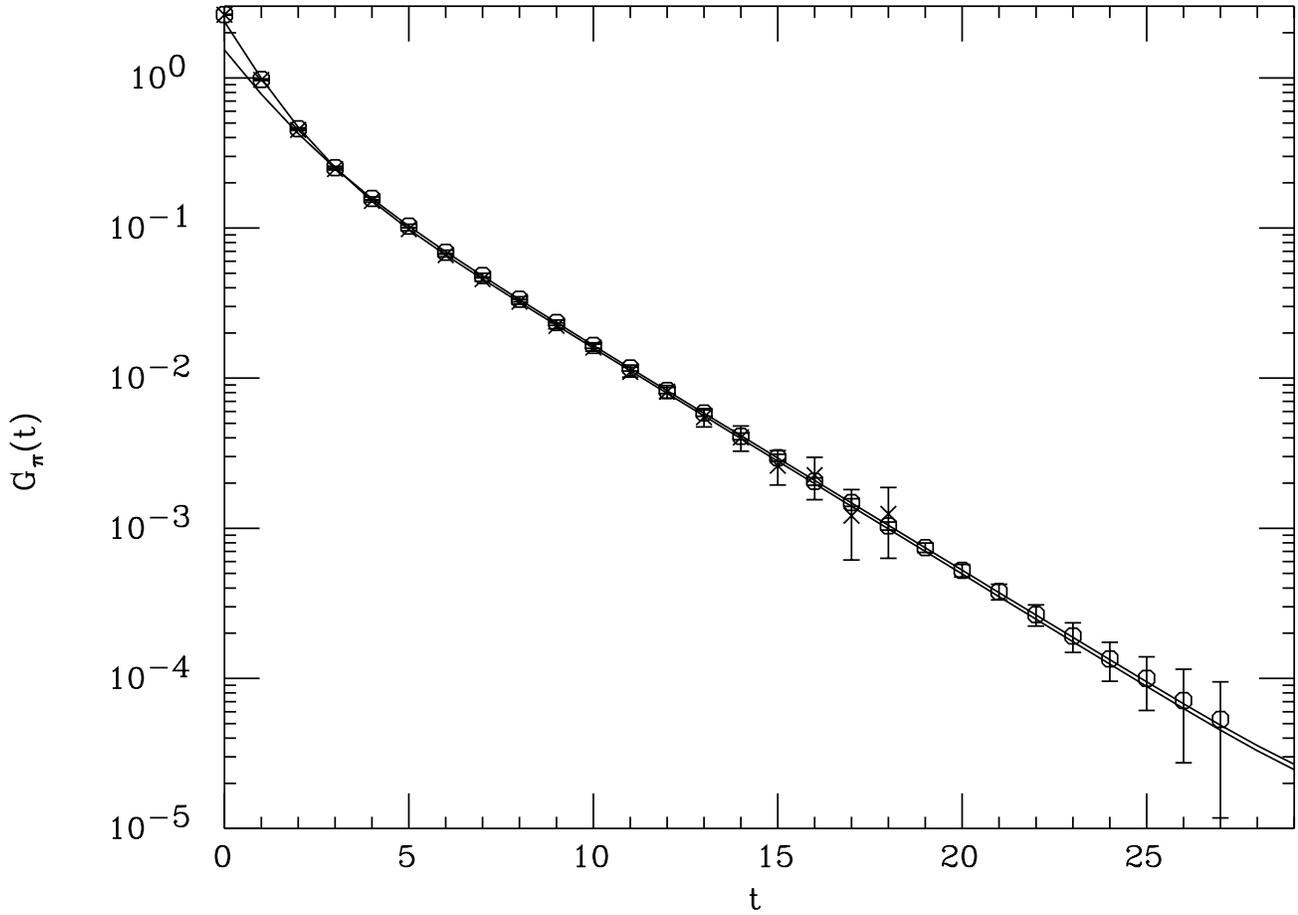

Figure 7. Two particle fits to the pion propagator in the time interval [5–32], at $\mu = .1$ (circles) and .15 (crosses). The fitting curves are practically coincident. Only the data points statistically different from zero are shown.

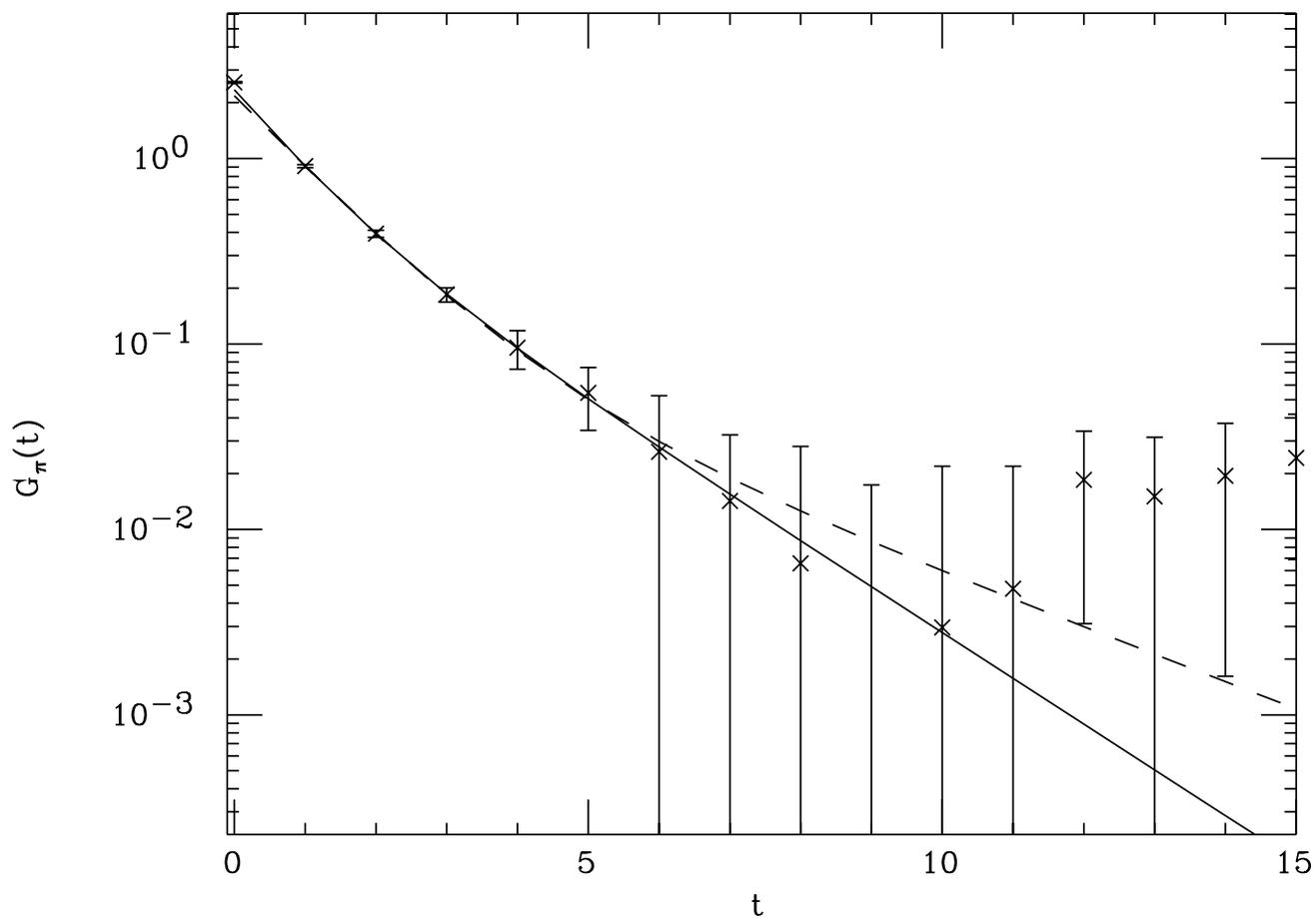

Figure 8. Two particle fit to the pion propagator at $\mu = .2$ in the interval [0–9] (solid line). A similar fit, but with the mass in the fundamental channel constrained to its $\mu = 0$ value is shown as a dashed line.

To get mass estimates at $\mu = .17$ and $\mu = .2$ we first restricted the fitted interval to a statistically significant range ($1 < t < 10$). Then we estimated the error due to the lack of asymptotic behavior by comparing, at $\mu = 0$, the result obtained on that range with the correct one, and we assumed that such an estimate is $\mu$ independent. In this way we estimated the mass and errors at $\mu = .17$, given in Table III, the first error being statistical and the second computed as just said. It is clear that this mass is certainly not asymptotic, and has to be considered only 'indicative'. The same problems occur at $\mu = .2$, where we have also tried some fits which discard the noisiest propagators. The situation at $\mu = .2$ is best described by a figure. We show in Fig. 8 two fits to the pion propagator at $\mu = .2$. The solid line is drawn in correspondence to the unconstrained fit, which gives for the central mass $m_\pi = .60$, with a statistical error $\Delta(m_\pi) = .24$. The dashed line is for a fit with a constrained value in the fundamental channel, $m_\pi = .3379$ as in the $\mu = 0.0$ case. The two fits are coincident in the region where the propagator is statistically different from zero. Only better results at large $t$ would allow a safe estimate for the pion mass.

### C. Further search for finite temperature effects

We have said that the results for the chiral condensate and the number density do not exclude the possibility of finite temperature effects. We have then selected three $\mu$ values, $.0, .15, .20$, where we had the same number of gauge configurations on the two lattices in order to do a more detailed comparison.

In Table V we record the results for the chiral condensate for the different boundary conditions, and opposite $\mu$ signs, in order to find systematic trends in their average values, or in their fluctuations. $\mu = 0$ sets the scale for the expected effects. We do not see any size effects at $\mu = 0$, while the error ratio is roughly $\sqrt{2}$, expected on a purely statistical basis (the lattice volumes differ by a factor of 2). Deviations from this trend have to be interpreted as induced by the chemical potential. At $\mu = .20$ it appears that the difference in the results is systematic, and statistically significant.

| | $< \bar{\psi}\psi >$ | | | | | |
|---|---|---|---|---|---|---|
| | $16^3 \times 32$ | | | $16^3 \times 64$ | | |
| $\mu$ | $Z_1$ | $Z_2$ | $Z_3$ | $Z_1$ | $Z_2$ | $Z_3$ |
| 0.0 | .1377(6) | .1377(9) | .1370(9) | .1376(4) | .1373(3) | .1379(4) |
| +.15 | .1371(16) | .1381(19) | .1368(15) | .1385(10) | .1390(8) | .1391(8) |
| −.15 | .1381(14) | .1317(69) | .1354(12) | .1377(10) | .1363(7) | .1381(9) |
| +.20 | .1145(90) | .1230(154) | .0843(248) | .1301(48) | .1328(82) | .1283(78) |
| −.20 | .1292(55) | .1073(142) | .0944(267) | .1262(72) | .1250(43) | .1237(36) |

TABLE V. Temperature dependence of $< \bar{\psi}\psi >$. We compare the results for the $16^3 \times 32$ and $16^3 \times 64$ lattices at $\mu = 0, \pm.15, \pm.20$, for the three different boundary conditions.

| | $< J_0 >$ | | | | | |
|---|---|---|---|---|---|---|
| | $16^3 \times 32$ | | | $16^3 \times 64$ | | |
| $\mu$ | $Z_1$ | $Z_2$ | $Z_3$ | $Z_1$ | $Z_2$ | $Z_3$ |
| 0.0 | -.00048(99) | .00081(117) | .00063(128) | .00034(88) | .00039(88) | .00015(90) |
| +.15 | -.00040(123) | .00228(297) | -.00219(140) | .00024(91) | .00170(72) | .00081(101) |
| −.15 | -.00232(123) | .00291(297) | .00078(161) | .00194(98) | .00053(83) | .00107(95) |
| +.20 | -.01665(973) | -.00311(1350) | .04192(3721) | .00548(499) | .00901(585) | .00819(541) |
| −.20 | -.01253(736) | -.00023(731) | .00433(1263) | -.00886(447) | -.00207(777) | .00477(324) |

TABLE VI. Temperature dependence of $J_0$. We compare the results for the two lattices at $\mu = 0, \pm.15, \pm.20$, for the three different boundary conditions.

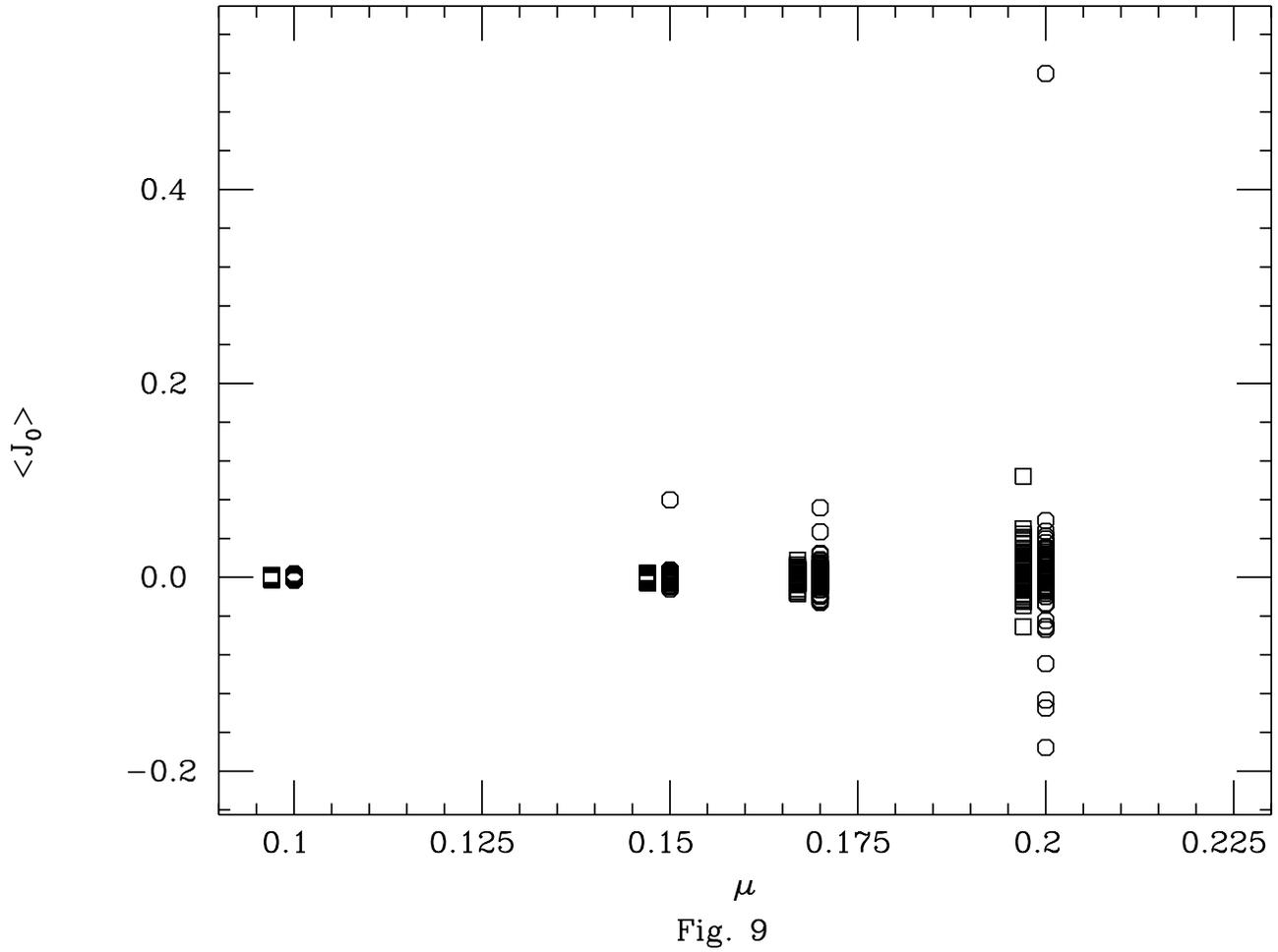

Figure 9. Collection of the results for the number density at $\beta = 6.0$, for the three different boundaries, as a function of $\mu$. The results for $\mu$ and $-\mu$ have been averaged. The circles are for the small lattice, the squares, displaced, for the big one.

Also, the error ratio grows larger, thus indicating some non–trivial improvement mechanism on the larger lattice, probably induced by the suppression of winding loops. Note also that the statistical gain coming from averaging over the boundary conditions is bigger on the smaller lattice, and that on the smaller lattice there is a much bigger dispersion from one boundary condition to another. This suggests that winding loops are playing a significant role, which diminishes as the lattice size increases.

Table VI, where the results for the number density are given, shows that $J_0$ is always consistent with 0. From Fig. 9 we see that on the smaller lattice (circles) there are exceptional events (deconfined configurations) which are absent on the bigger lattice (squares). Note that some of the exceptional configurations have a density of the "wrong" sign: for instance, a negative chemical potential is apparently inducing a positive density!

In conclusion, by increasing the temporal extent, lattice artifacts are lessened. However, the improvements are very modest, so there is no reason to believe that the problems inherent in quenched simulations will be solved simply by using large volumes.

## III. THE INFINITE COUPLING LIMIT OF NON–ZERO DENSITY QCD

Recall, once more, the nature of the finite-density quenched QCD puzzle : observables deviate from their $\mu = 0.0$ values when $\mu = m_\pi/2$. In particular, the chiral condensate appears to fall at $\mu = m_\pi/2$. This result, which has suggested to other workers in the field that an infinitesimal value of the chemical potential restores chiral symmetry in the chiral limit, has been observed in simulations in the scaling region, as well as in the strong coupling limit [4]. At strong coupling this onset of sensitivity to $\mu$ observed in numerical simulations also differs from analytic predictions. This is rather peculiar, since analytic mean-field strong-coupling results usually compare well with numerical simulations. It will prove to be informative to uncover the reasons behind the discrepancy at strong coupling because it will suggest the reason for the failure of the quenched approximation in general. These considerations, together with the low computational costs of these simulations (compared to the simulations in the scaling region described above), motivated the following study.

We generated 20 random configurations on a $8^3 \times 16$ lattice, and 10 configurations on a $8^3 \times 32$ lattice. Again, we used three different anti-periodic boundary conditions for the fermions in the time directions : $\psi_i(t) = -Z_i\psi(0)$, where $Z_i$ stands for each of the three cube roots of unity. We first chose a bare quark mass of .1. The strong coupling predictions for the pion and baryon masses at $m_q = .1$ are .6 and 3.3, respectively. In our simulations $\mu$ ranged from 0. to 1.2, thus including the interesting region $.3 < \mu < 1.1$ and the mean field prediction for the pseudo-critical $\mu$ ( $\mu_c \simeq 0.6$). Moreover, a subset of 7 configurations on the smaller lattice was analyzed with a very heavy quark mass ($m_q = 1.5$), with $\mu$ in the interval (1.2:1.5), the pseudocritical point in this case being $\mu_c \simeq 1.37$. Finally, 10 configurations on an $8^3 \times 32$ lattice with $m_q = .1$ were used to monitor possible temperature effects.

We have measured the chiral condensate, the number density, the energy density and the pion mass. The other masses are too heavy, and difficult to extract – a typical limitation of strong coupling simulations. Along with standard observables, the unphysical operator $G_\mu G_\mu^\dagger$ (the "false" or "baryonic" pion, suggested by [6] as opposed to the real pion $G_\mu G_{-\mu}^\dagger$) was measured to obtain information about the poles of the Dirac operator.

The main results of this study are :

First, there are three distinct intervals of chemical potential. A conventional region, for $\mu < m_\pi/2$, where all the zero–triality observables maintain their $\mu = 0$ value. A "forbidden" region, for $m_\pi/2 < \mu < m_N/3$ (note the dependence on the nucleon mass), characterized by wild fluctuations, and by average values which (when statistically meaningful) show deviations from their $\mu = 0.0$ values. Finally, a saturation region, where all the thermodynamic observables have their limiting, large $\mu$ values, the amplitudes in the mass spectrum drop to zero, and no single particle states exist.

Moreover, inspection of the effective potential as computed in a mean field approximation shows that the "forbidden" region, which begins at $\mu = m_\pi/2$, corresponds to the onset of metastability in the effective potential.

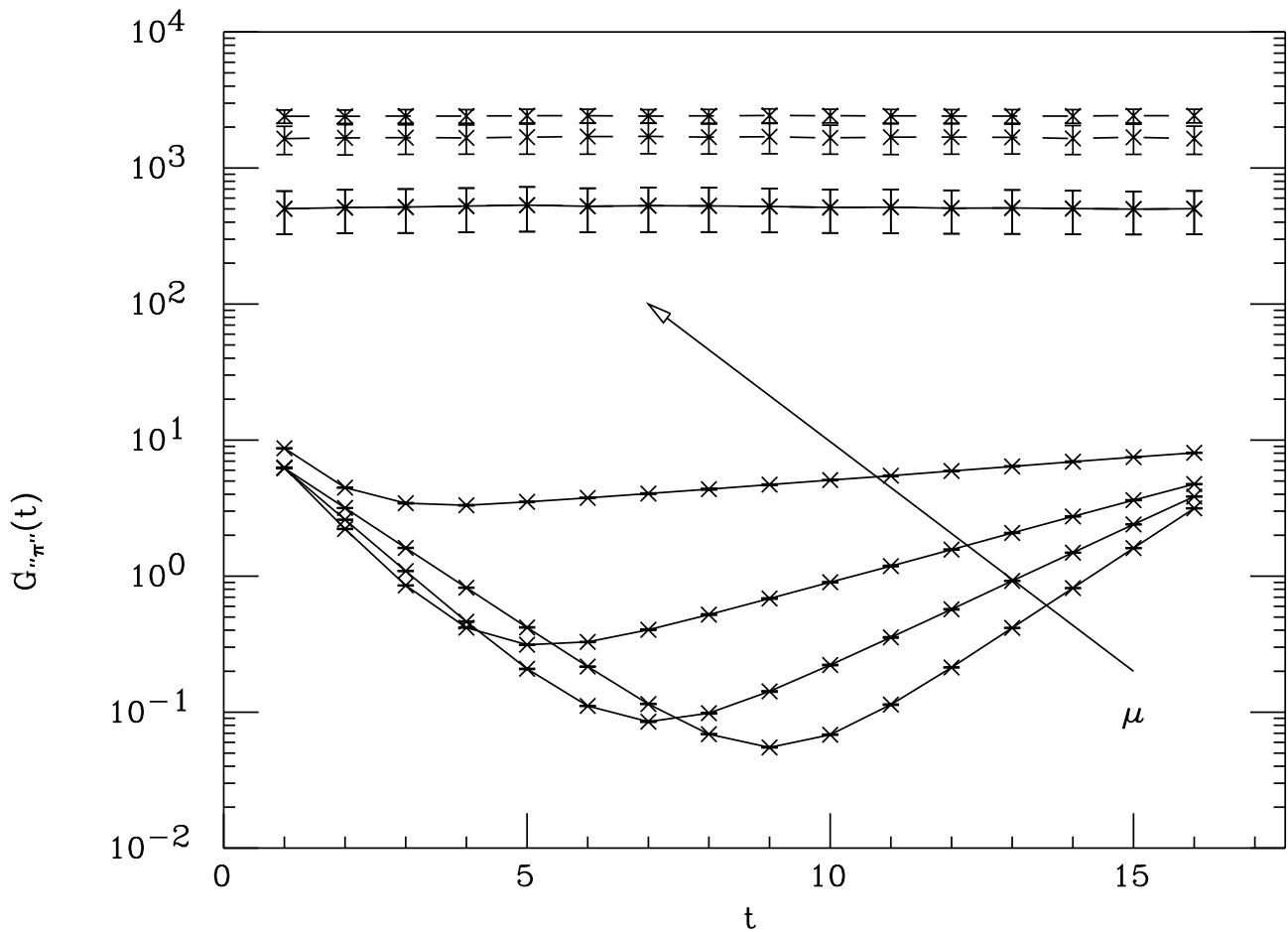

Fig. 10

Figure 10. The propagator of the "baryonic" pion is shown as a function of time for $\mu =$ 0.0, .1, .2, .3, .4, .5, .9. Note the T-asymmetry for $\mu < m_\pi/2$, and the flatness associated with the forbidden region.

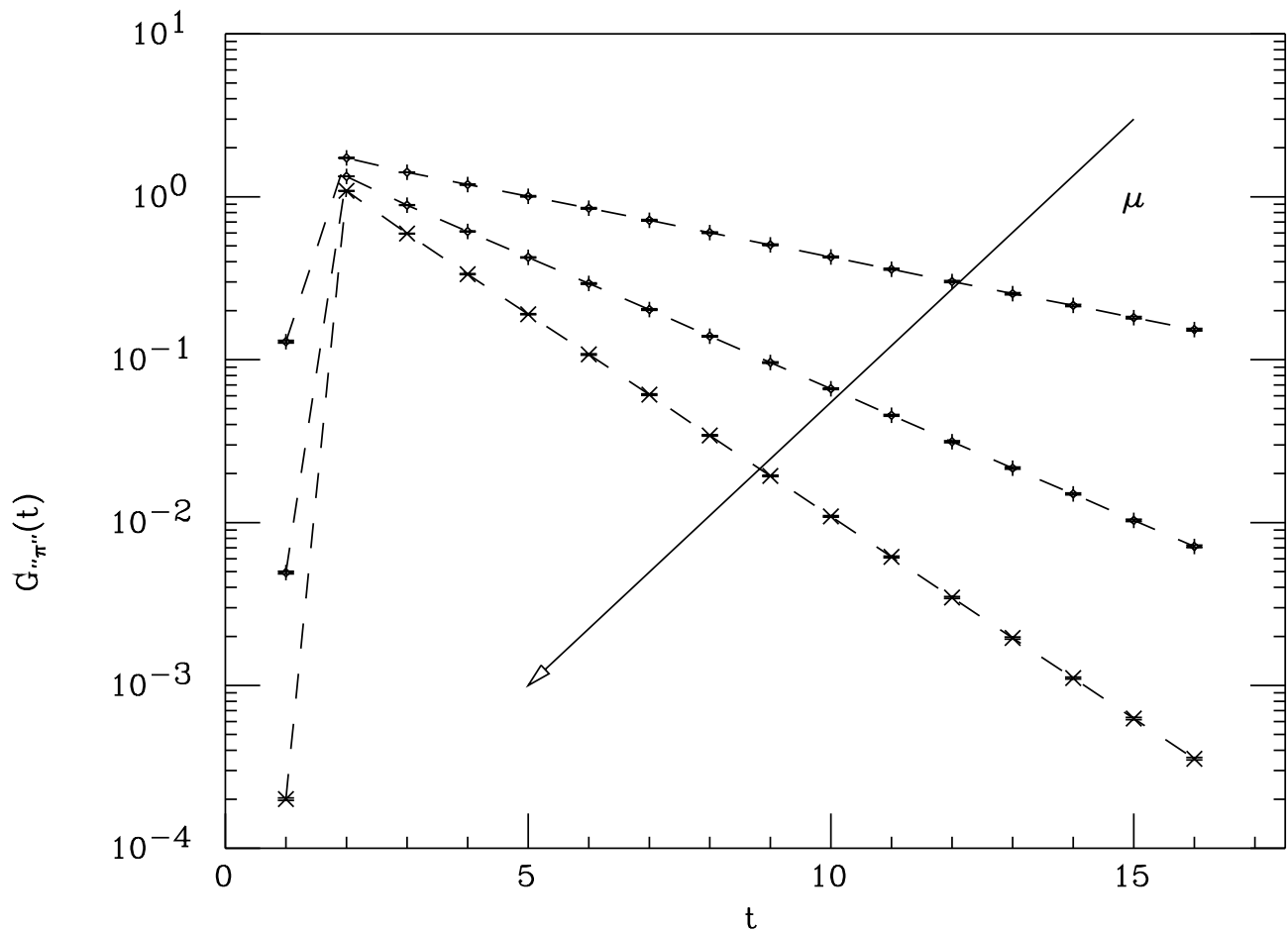

Figure 11. The propagator of the "baryonic" pion for $\mu = 1.0, 1.1, 1.2$, in the saturation region.

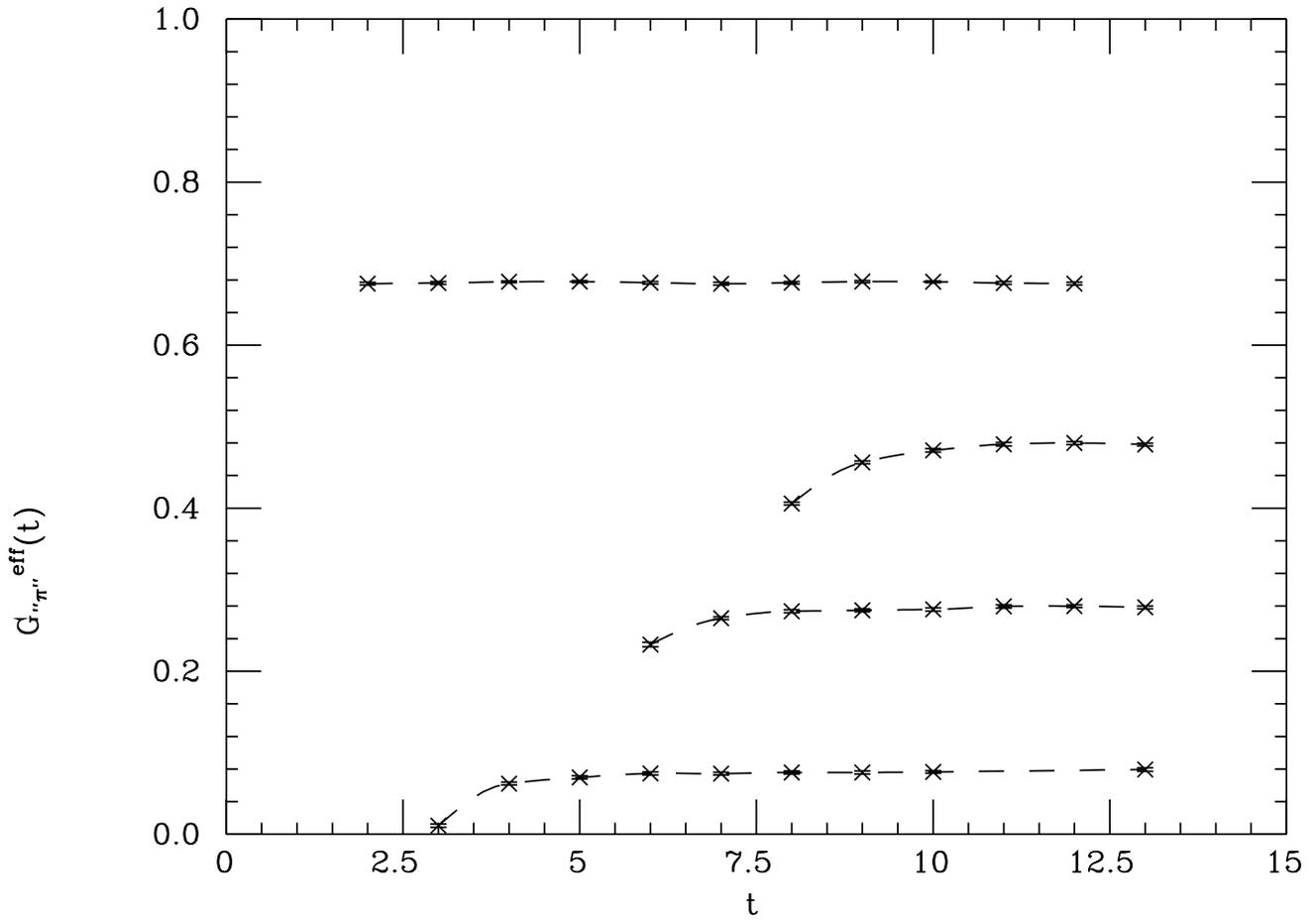

Figure 12. "Pion" effective mass versus time, at $\mu = (0.0, .1, .2, .3)$, from top to bottom

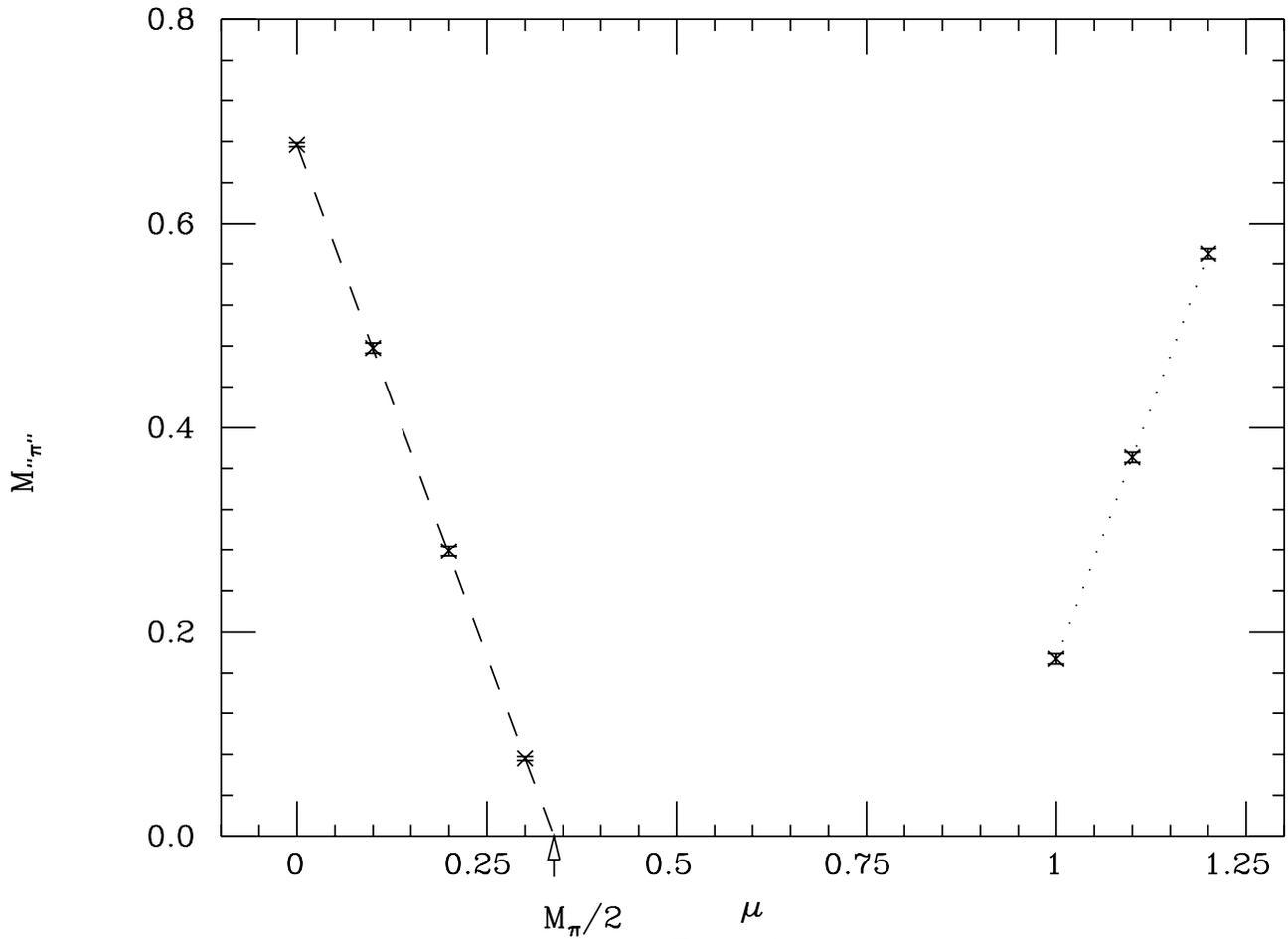

Figure 13. "Pion" masses, from an effective mass analysis, versus the chemical potential $\mu$. The dashed line is $y = m_\pi(0) - 2\mu$.

## A. Fluctuations, poles of the Dirac operator and the "baryonic" pion

As discussed at length in past work, when the chemical potential $\mu$ exceeds half the pion mass, the fluctuations increase dramatically. This can be read off the eigenvalue distribution [2], since when $\mu \simeq m_\pi/2$ the quark mass lies inside the eigenvalue cloud, thus producing eigenvalues of the full Dirac operator close to zero. These huge fluctuations are associated with a dramatic increase in the CPU time required for the inversion of the Dirac operator: the average number of conjugate gradient iterations for $\mu < m_\pi/2$ was $\simeq 100$, while when $\mu$ exceeds $m_\pi/2$ the number of iterations grows large ($\simeq 10^4$). Again, this is consistent with the eigenvalue picture, the number of iterations being proportional to the ratio of the largest to the smallest eigenvalue. For $\mu \gtrsim 1$ the number of iterations is again very small ($\simeq 10$). In this region we are inside the eigenvalue "crescent", and the first Brillouin zone is saturated. Again, we note that this value corresponds, approximatively, to one third the baryon mass: the theory, at least in its strong coupling limit, is sensitive to the nucleon, as predicted by simple nuclear models.

In order to gain additional insight into the eigenvalue distribution and zero modes of the Dirac operator, we studied a suitable, "unphysical" operator, the "baryonic pion" (which we shall often denote " $\pi$ "): $G_\mu G_\mu^\dagger$, as opposed to the real pion $G_\mu G_{-\mu}^\dagger$, $G_\mu$ being the quark propagator computed with chemical potential $\mu$. We shall see that the amplitude of the baryonic pion propagator grows very large when $\mu > m_\pi/2$, and its poles, which are related to the poles of the Dirac operator, are very well exposed.

We show the " $\pi$ " propagator in Figs. 10 and 11. For small values of the chemical potential the propagator has a baryon–like (T– asymmetric) behavior. For $\mu > m_\pi/2$ the propagator flattens, which can only happen because of zero eigenvalues of the Dirac operator: this gives direct evidence of the zero modes we were looking for (Fig. 10). In the saturation region (Fig. 11), $G_{"\pi"}(t)$ is again baryon–like, but this time the propagation, as expected, is in the opposite direction. Outside the forbidden region, which, in this case, is associated with the flat propagators, we can measure the baryonic pion mass. For that purpose, we relied on a simple effective mass analysis, and evaluated the logarithmic derivative without any attempt at symmetrization (we hesitated to use the same parameterization as for staggered baryons). The flat region in the effective mass plots extends till the influence of the backward propagating state becomes appreciable : from our results it is clear that this still produces a reasonable interval, so that we can obtain a safe estimate for the baryonic pion mass. In Fig. 12 we show the effective mass plots, together with the results of a conventional effective mass analysis for the real pion. The "$\pi$" mass satisfies, for $\mu < m_\pi/2$, $m_{"\pi"}(\mu) = m_{"\pi"}(0) - 2\mu$, (where $m_{"\pi"}(0) = m_\pi(0) = m_\pi$) thus extrapolating to zero for $\mu = m_\pi/2$ (Fig. 13).

In conclusion, the study of the baryonic pion gives clear evidence of zero modes in the quark propagator for $\mu \simeq m_\pi/2$, in agreement with the eigenvalue picture.

Summarizing, these measurements provide a coherent description of the spectral structure of the Dirac operator over the entire $\mu$ range. The poles for $\mu > m_\pi/2$ can be clearly exposed by suitable, yet unphysical observables. Their mathematical and physical significance is open to many interpretations which have been discussed and reviewed in the literature. We shall comment further on this point below.

### B. Physical observables

We have measured the particle spectrum, and thermodynamic observables (chiral condensate and number density), via a standard stochastic estimator.

The pion propagator (Fig. 14) is completely insensitive to the chemical potential up to $m_\pi/2$. The "forbidden" region ($m_\pi/2 < \mu < m_B/3$) is dominated by fluctuations, while in the saturation region ($\mu > m_B/3$) the pion propagator amplitude drops to zero. We were not able to measure the baryon mass (a common drawback of strong coupling simulations), thus missing a very important piece of information about the critical behavior. It was however possible to monitor the amplitude of the baryon propagator. At $m_q = .1$ we found that it decreased by an order of magnitude in the saturation region ( the amplitude is $\simeq .01$ in the conventional region, undefined in the forbidden one, and $10^{(-6,9,11)}$ at $\mu = (1, 1.1, 1.2)$). This trend in the amplitude is common to all the propagators we have measured, and indicates that no real particle can exist in the saturation region.

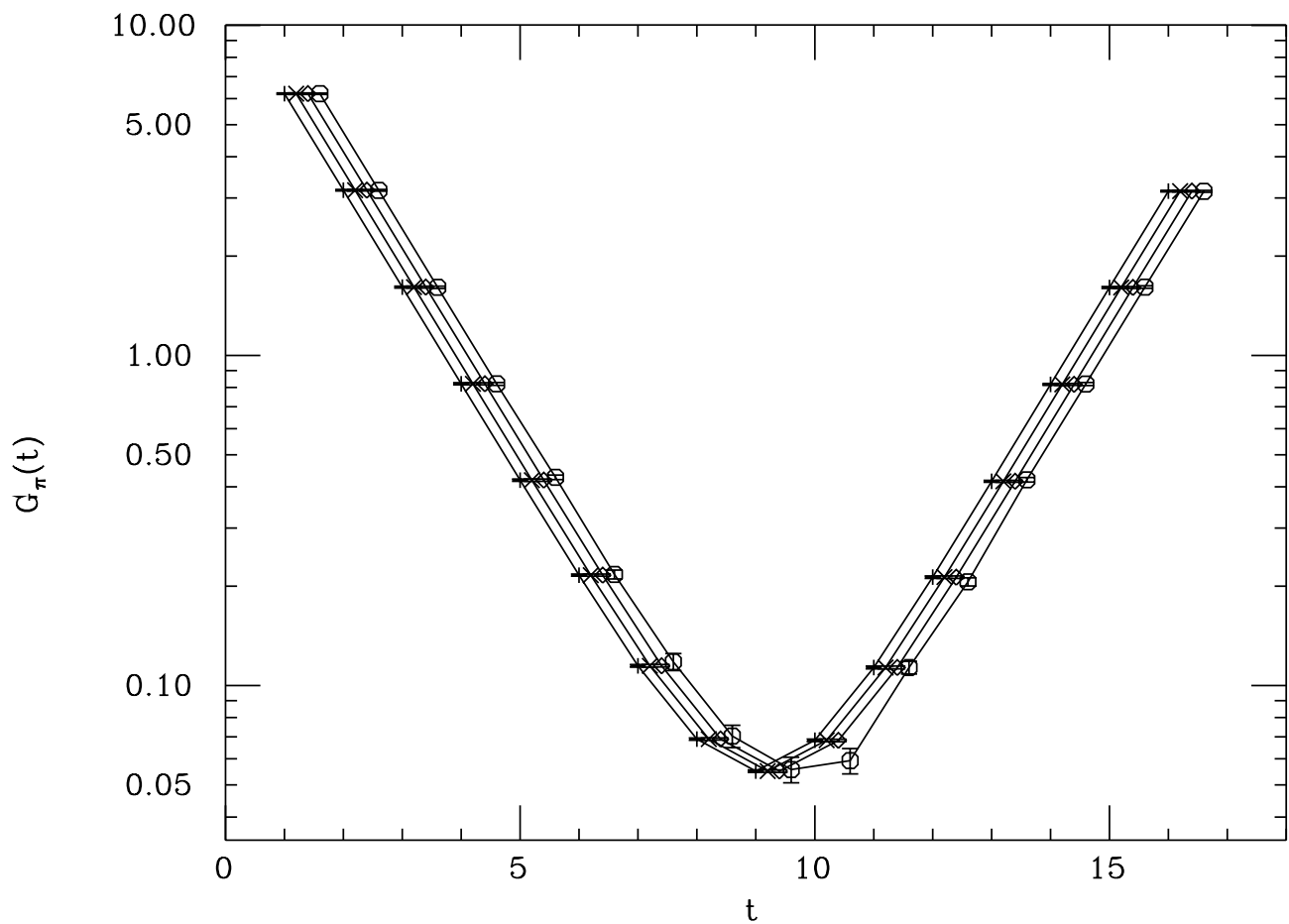

Fig. 14

Figure 14. Pion propagators at $\mu = (0.0, .1, .2, .3)$ (Plus's, crosses, diamonds, circles.) The pion mass is insensitive to the chemical potential in this range of values.

The results for the thermodynamic observables are shown in Figs. 15, for $m_q = .1$ (Fig. 15a) and $m_q = 1.5$ (Fig. 15b). As noticed in past work, the chiral condensate and the number density are closely correlated. Saturation is especially clear in the number density, which is three –the maximum number of quarks which can occupy a single site– for $\mu > m_N/3$. Our results are consistent with those previously reported in the literature, whenever a comparison is possible ($\mu < m_\pi/2$).

It is also informative to consider the results configuration by configuration. We know that ensemble averages have some special features at finite density, and a more detailed analysis can be worthwhile. For that purpose, we use the data at $m_q = .1$ which has the best statistics.

In Figs. 16 we show the number density as a function of the configuration number. At low $\mu$ (Fig. 16a), the densities computed with opposite values of the chemical potential (joined with dotted and dashed lines, respectively) are strongly correlated: so we do not see any sensitivity to the chemical potential itself. In the intermediate region (Fig. 16b), $J_0$ is wildly oscillating. We observe configurations which have a non–zero density, but with the wrong sign, for instance when a positive density is obtained with a negative $\mu$. This, once more, tells us that one must be careful in using concepts like confinement and chiral symmetry breaking on an isolated configuration: a negative chemical potential should enhance, in the statistical average, antibaryon propagation, while we see that on isolated configurations the opposite can happen. In the saturated region (Fig. 16c) things are very clear, and each configuration has a net density with the "right" sign.

In Figs. 17 and 18 we give the full ensemble of values for the chiral condensate and the number density, as a function of the chemical potential. For the number density, we took the average over $\pm\mu$, which enforces T-symmetry event by event, and produces $J_0 = 0$ at $\mu = 0$.

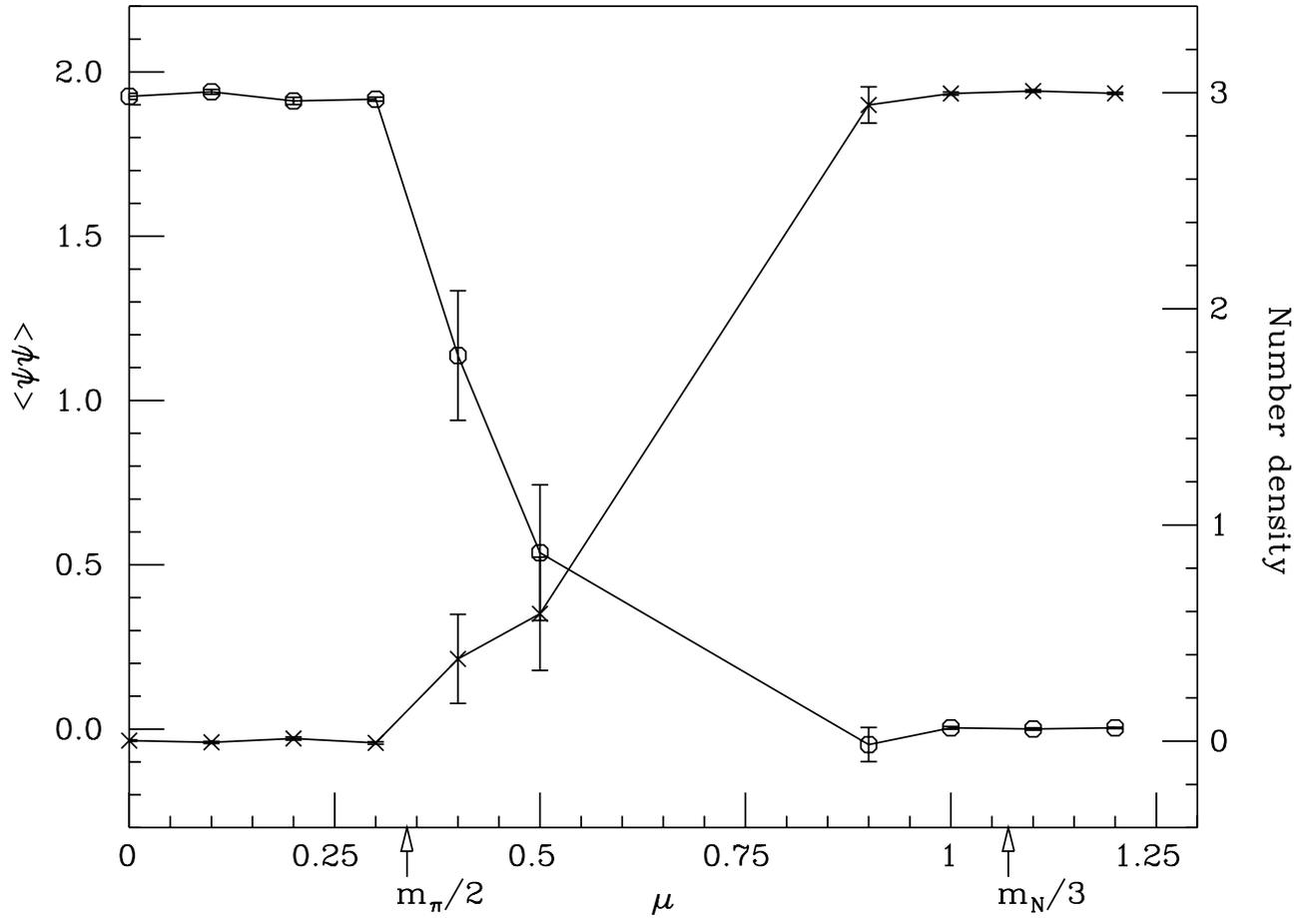

Figure 15a. Summary of the thermodynamics results at strong coupling. The chiral condensate ( circles, left ) and the number density ( crosses, right ) are plotted as a function of $\mu$ , for $m_q = .1$

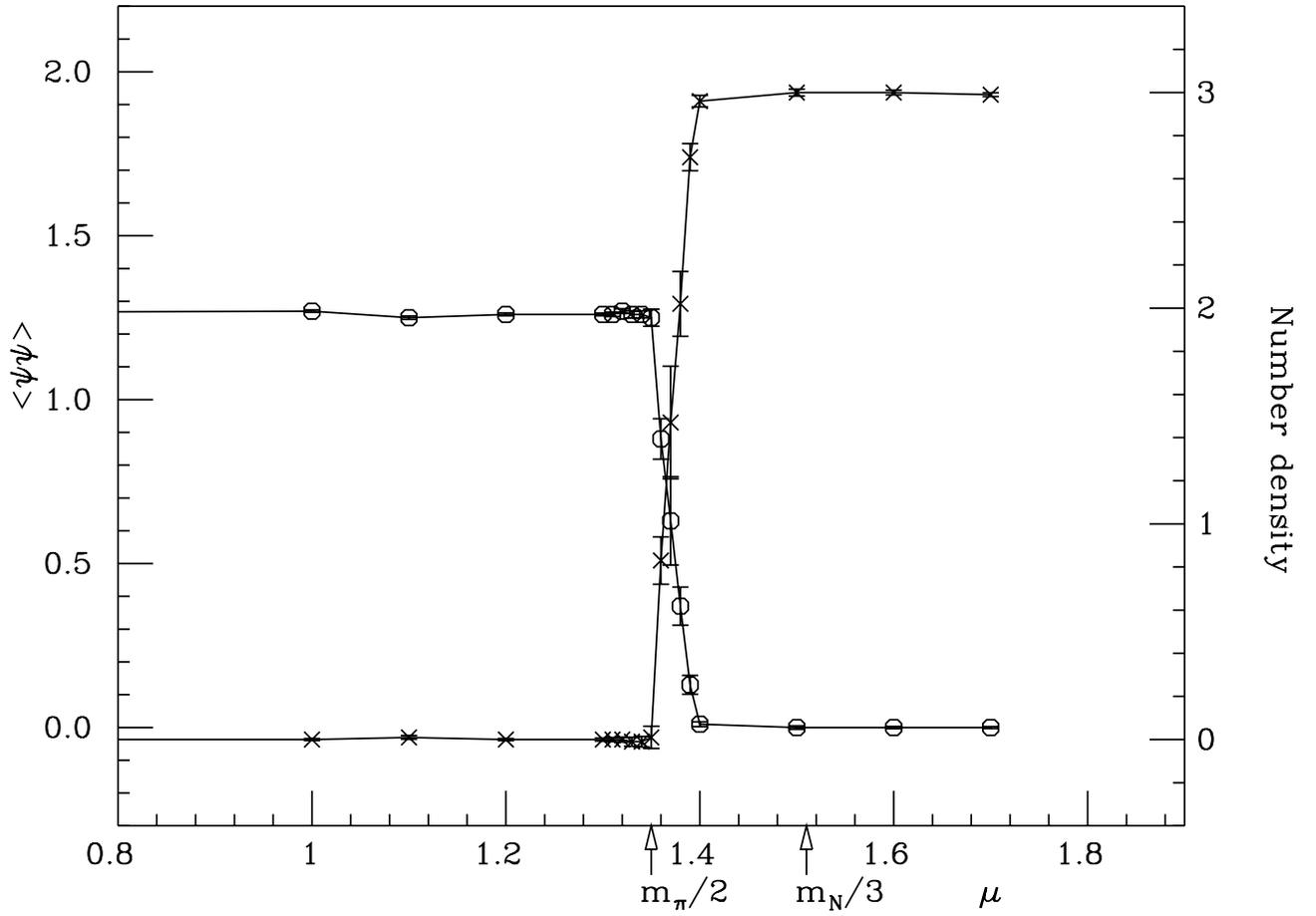

Figure 15b. As in Fig. 15a, but $m_q = 1.5$

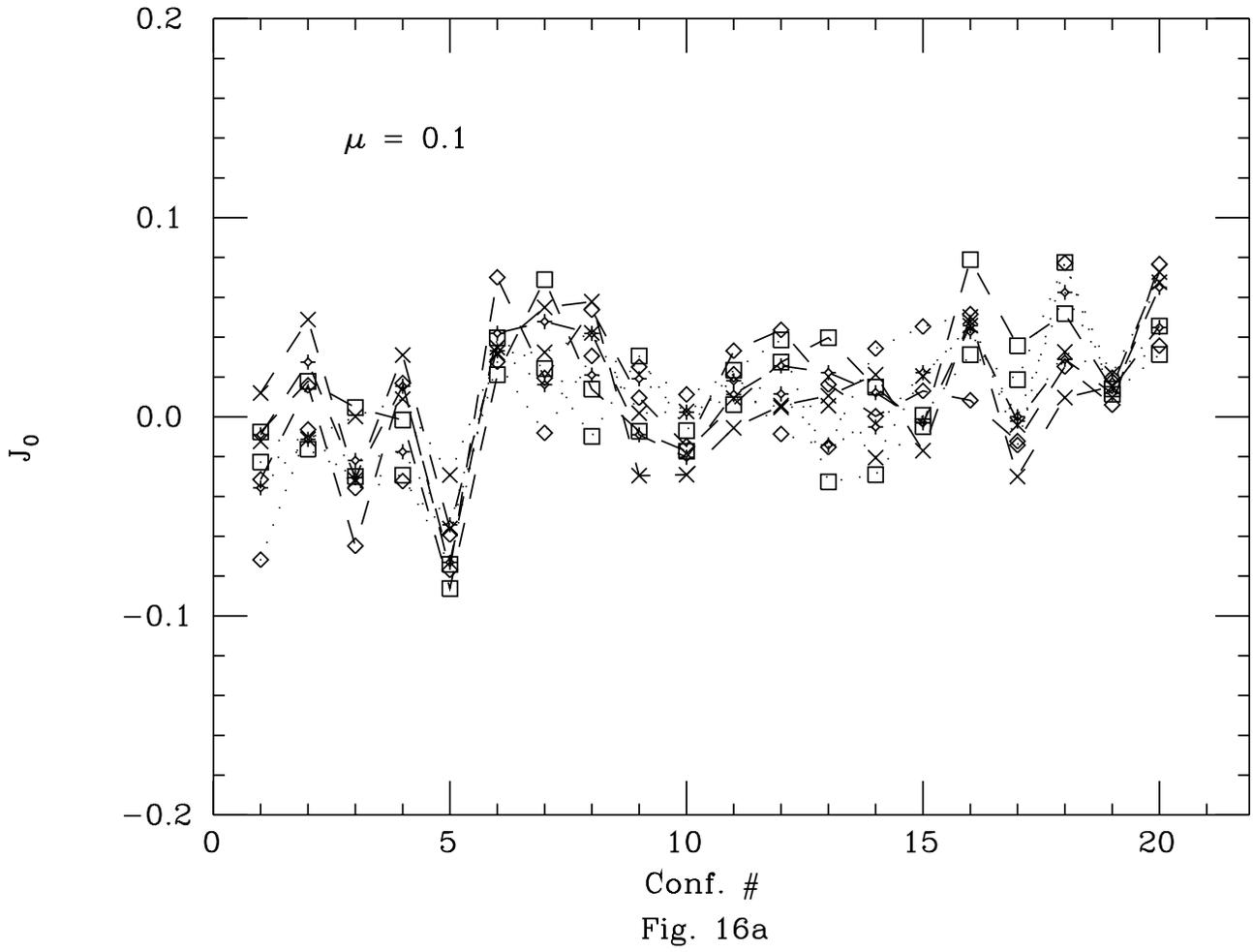

Figure 16a. The results for the number density $J0$ are shown configuration by configuration, for the three boundaries and the two opposite $\mu$ values at $\mu = .1$

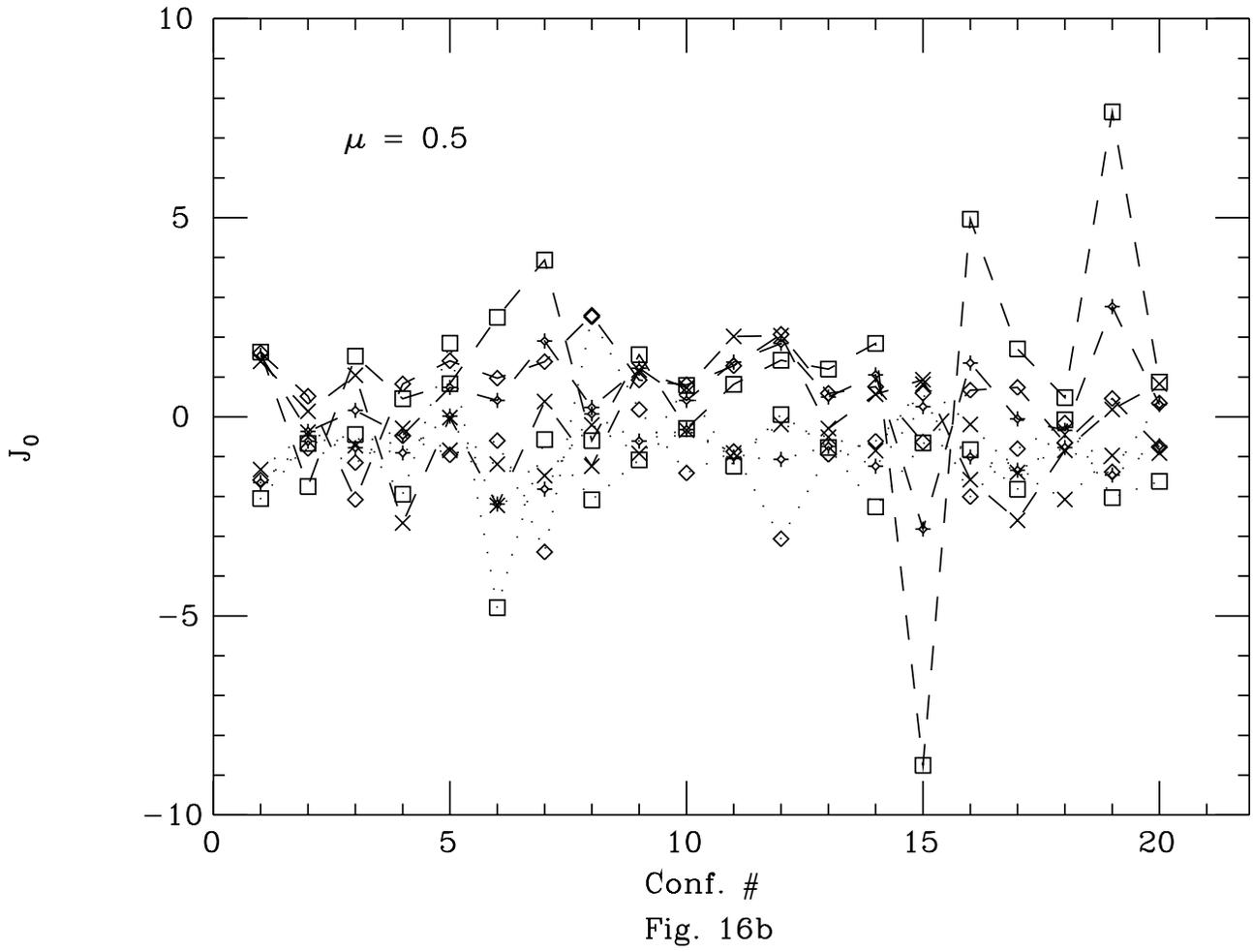

Figure 16b. As in Fig. 16a, but $\mu = .5$

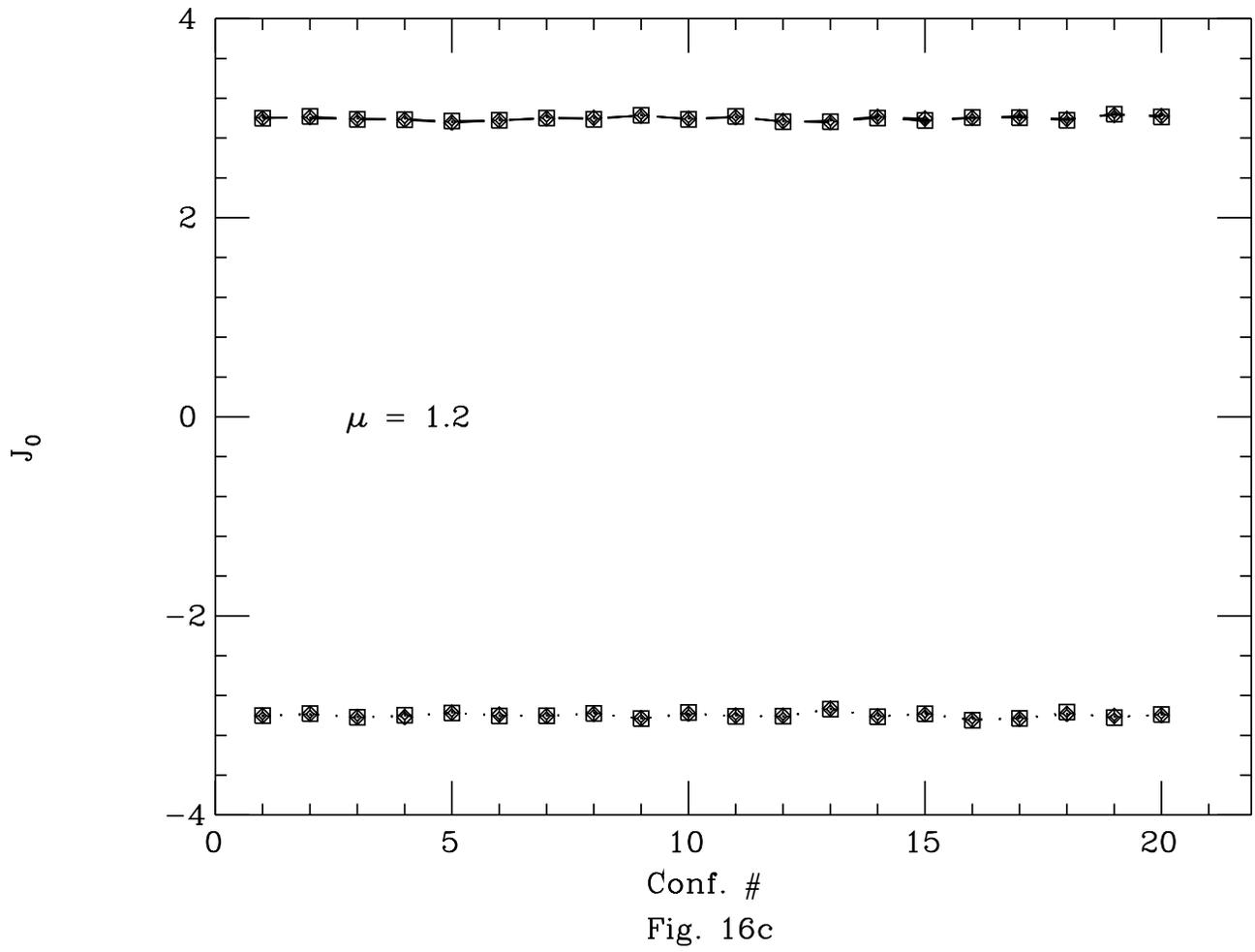

Figure 16c. As in Fig. 16a, but $\mu = 1.2$

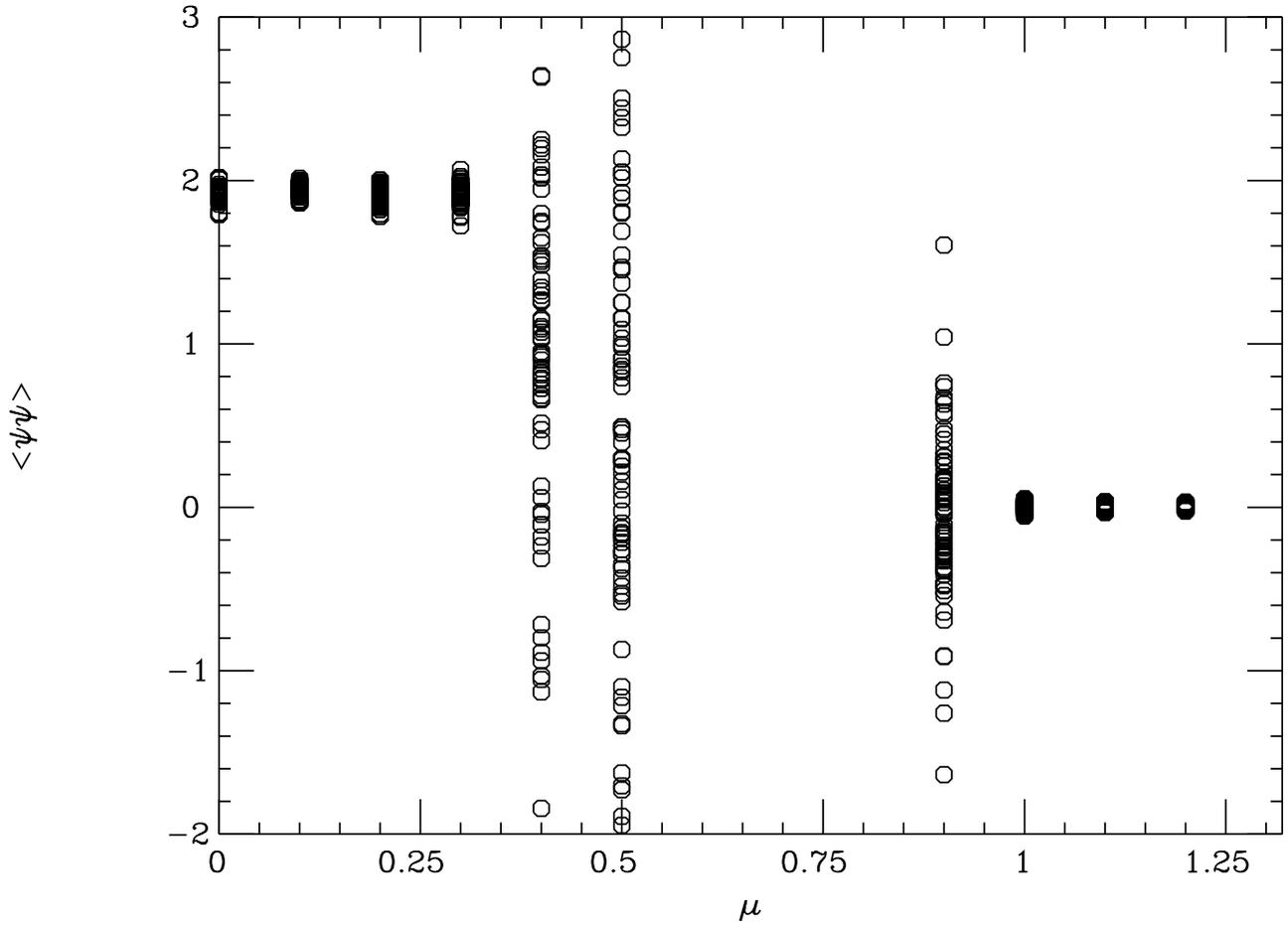

Figure 17. Collection of the results for the chiral condensate, for the three different boundaries, and the opposite values of $\mu$, as a function of the chemical potential.

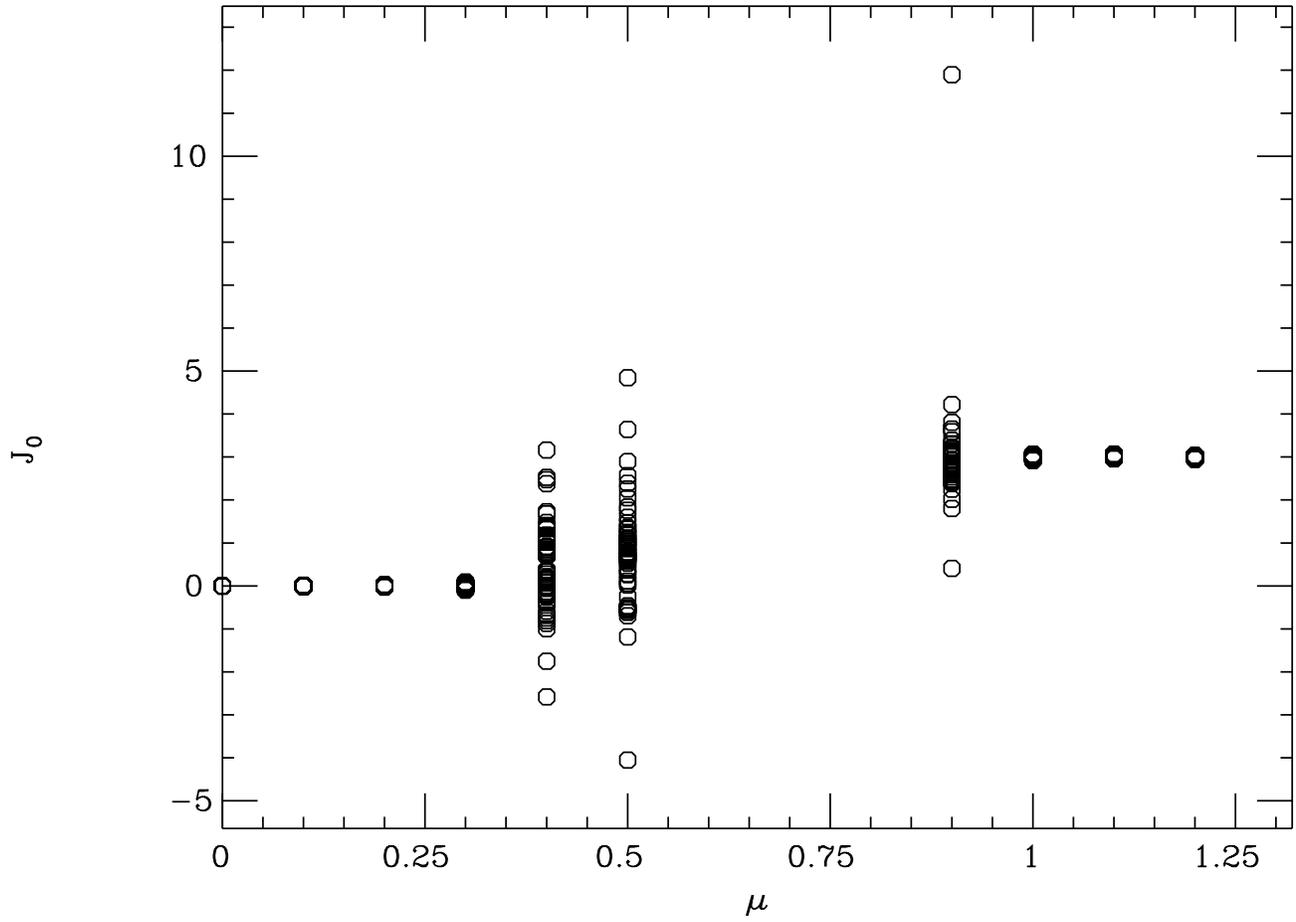

Figure 18. Collection of the results for the number density, for the three different boundaries, as a function of $\mu$. The results for $\mu$ and $-\mu$ have been averaged

For $\mu > m_\pi/2$ the results for $<\bar{\psi}\psi>$ and $J_0$ obtained configuration by configuration spread over a wide range, which includes both the saturation and the $\mu = 0$ values.

The behavior we have described suggests that we might have entered a metastable region rather than the symmetric phase, and that the identification of the onset of $\mu$ dependence in the theory's observables with the pseudocritical point computed in a mean field approximation may not be correct. In the next subsection we shall compare this interpretation of the data with the predictions of the strong coupling analytic calculations.

### C. Effective models and simulation results

Several, qualitatively equivalent, strong coupling, $1/d$ expansions of full QCD with staggered fermions at finite density and temperature are available in the literature [4] [7] [8] [10]. Such mean field analyses predict a strong first order transition at finite chemical potential and zero temperature. Since fermion loops do not play a significant role in the leading order of the $1/d$ expansion at strong coupling, it is sensible to try to interpret quenched simulations in terms of the analytic mean field predictions. This has been done in the past [4] [7], and we will continue in this tradition in some detail. We shall show that the onset $\bar{\mu}$ of the metastable region of the effective potential of these studies is associated with the onset $\mu_{onset}$ for $\mu$ dependent thermodynamics introduced at the beginning of this paper.

Let us now describe these calculations in more detail. At small $\mu$ the free energy, which plays the role of an effective potential, plotted as a function of $\bar{\psi}\psi$ has only one minimum, corresponding to the $\mu = 0.0$ value of the chiral condensate $<\bar{\psi}\psi>_0$. While increasing $\mu$ = $\bar{\mu}$, the free energy develops a secondary minimum at $\bar{\psi}\psi = 0$. The two minima become of equal height for $\mu = \mu^*$. $\mu^*$ is thus interpreted as the pseudocritical point at finite mass, or the exact critical point $\mu_c$ for chiral symmetry restoration in the chiral limit. $(\bar{\mu} - \mu^*)$ approaches 0 as the quark mass increases, while these two values of the chemical potential are well separated for smaller quark masses, thus creating a wide metastable region. By further increasing the chemical potential, the minimum at finite $\bar{\psi}\psi$ eventually disappears at $\mu = \mu_s$, the only surviving minimum being the one at $\bar{\psi}\psi \simeq 0$. This should correspond to the complete saturation of the first Briouillin zone, i.e. density = 3. Before proceeding, we recall that $\mu_c \neq m_B/3$. This can be understood because of a considerable nuclear binding energy at strong coupling [7] (infact, $\mu_c$ approaches $m_B/3$ when $g \to 0$), or because of the effects suggested in Ref. [16]. Anyway, from the perspective of the present study, the main feature of the strong coupling analysis is that the critical chemical potential $\mu_c$ is different from zero, and the lack of exact coincidence with the naive predictions $m_B/3$ carries little weight: if the results of the strong coupling analysis were realistic, the theory would not be pathological. This is why it is important to reconcile the numerical and analytic results at finite mass, or to uncover the reasons behind their differences.

Summarizing, there are a few interesting values of the chemical potential, labeling the different physical regions associated with the chiral transition which emerge from the effective potential analysis: $\mu_c$, the critical point of the chiral transition in the chiral limit; $\mu^*$, the pseudocritical point as analytically computed at finite mass; $\bar{\mu}$, the onset of the metastable region found in an effective potential approach; $\mu_s$ the chemical potential at which saturation occurs. We would like to study the interrelation of these points with $\mu_{onset}$, the onset of $\mu$ dependence in thermodynamic observables at finite quark mass observed in numerical simulations. Naive arguments would suggest that $\mu^* = \mu_{onset}$, i.e. that the transition has a strong first order character also at finite mass. However, it is easy to imagine situations (as we are will explain below) in which $\mu^*$ and $\mu_{onset}$ are different. In this case, $\lim_{m_q \to 0} \mu^* = \mu_c$, while $\mu_{onset}$ in itself does not have any immediate relationship with the critical point $\mu_c$. Rather, it should be identified with $\bar{\mu}$ : the onset of $\mu$ dependence in the thermodynamic observables is produced by the edge of the metastability region. We also mention that the work of Ref. [13] supports the possibility that $\mu_{onset}$ and $\mu_c$ are indeed different. In Ref. [13] a representation of the partition function at strong coupling written in term of a monomer–dimer expansion is used to compute the order parameter and the number density. $< \bar{\psi}\psi >$ deviates from its $\mu = 0$ value at the value $\mu^*$ of the chemical potential computed in a mean field approximation. However the onset $\mu_{onset}$ for $\mu$ dependence in thermodynamic quantities, as inferred from the behavior of the number density, occurs much earlier.

We now turn to a discussion of the results: we shall compare the results for the effective potential drawn in Figs. 19a and 19b with the thermodynamic results presented in Figs. 15a and 15b.

In the following, we use results and notation from Ref. [7]. We show in Fig. 19a the effective potential (after formulæ(2.10 and segg. of Ref. [7] ) for the same $\mu$ values of our simulations at $m_q = .1$, the same quark mass, the asymmetry parameter $a_\tau/a_s = r = 1$, and $T = 16$. Also, we show as a dotted line the effective potential at $\mu = .7 \simeq \mu*$. The curves are basically coincident for "large" $\bar\psi\psi$, while for small values they are qualitatively different. In particular, we clearly see the onset of metastabilities at $\bar\mu \simeq .4$, and the disappearance of the minimum at finite $\bar\psi\psi$ for $\mu \simeq 1$. The pseudocritical point $\mu^*$ is obscured by the huge fluctuations associated with such metastabilities. It is only when $\mu > m_B/3 \simeq \mu_s$ that the minimum at finite $\bar\psi\psi$ disappears, and for $\mu > m_B/3$ all the observables have their limiting values: the chiral condensate is zero and the number density is three. By contrasting this behavior with Fig. 15a we see indeed that $\mu_o \simeq \bar\mu$, and that saturation occurs for $\mu > \mu_s$. In Fig. 19b we have repeated this plotting exercise for our higher mass value ($m_q = 1.5$), with an analogous result. In fact, we have observed that in this case the metastable region ($\mu_o < \mu < \mu_s$) in the effective potential shrinks to the expected small interval, as predicted from Fig. 15b.

We learn that the metastable region in the effective potential can be identified with the forbidden region observed in the simulations. This clarifies the nature of the phenomena observed at $m_\pi/2$ : the pathologies are not manifestations of an early chiral transition, but are indicative of the existence of a region of metastability. For $\mu \simeq m_\pi/2$ we have the first appearance of zero modes. These manifest themselves in huge fluctuations, in the increase of the iterations needed for the inversion, and in the flatness of the baryonic pion propagator. In particular, the fluctuations observed in the chiral condensate should be associated with fluctuations and distortions of the eigenvalue distribution: when the chiral condensate is almost zero, the eigenvalue distribution must be such that the point $(m_q, 0)$ gets inside.

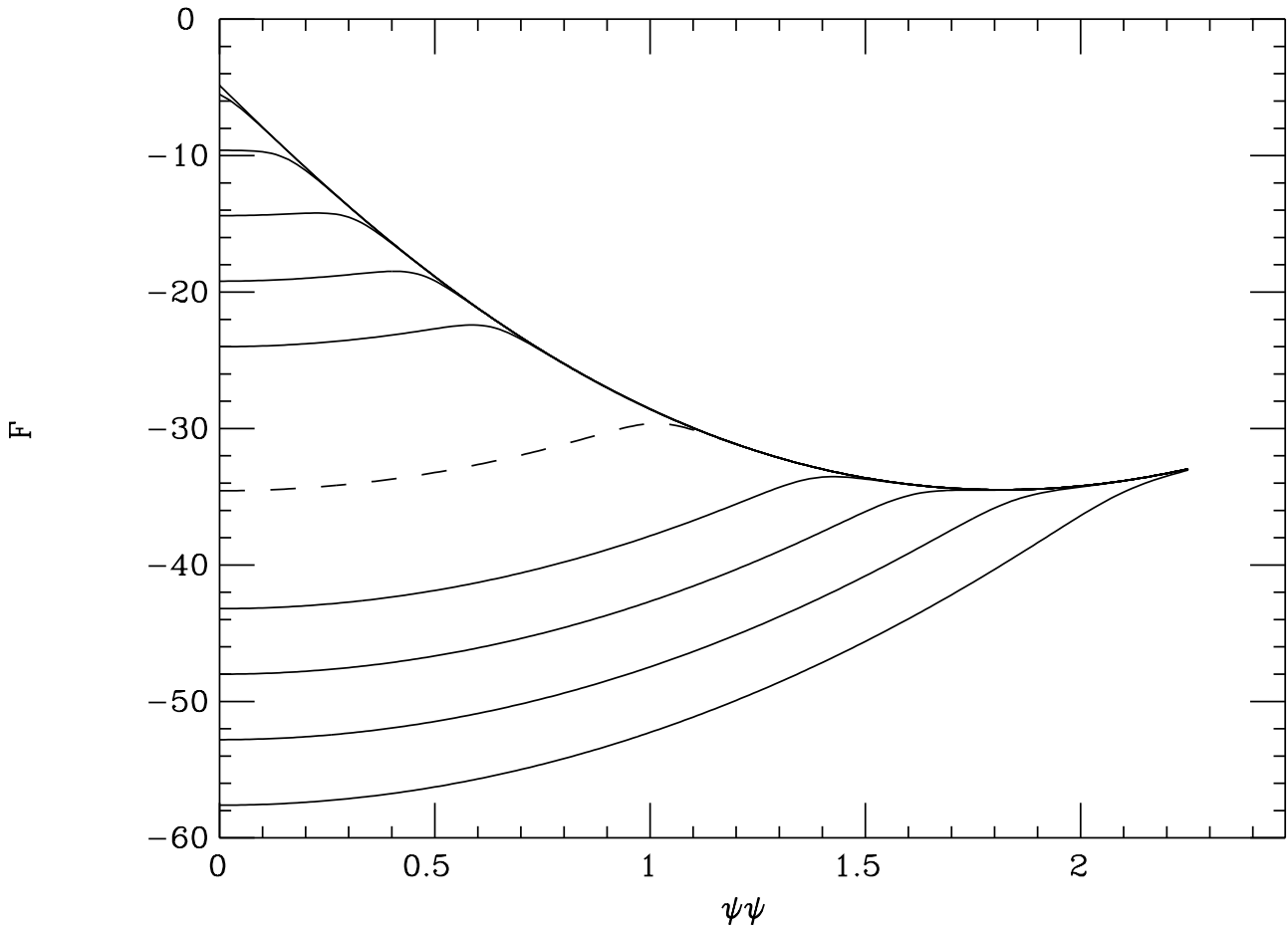

Figure 19a. Free energy, from Ref.7, as a function of $\bar{\psi}\psi$, for $m_q = .1$ and the chemical potentials used in our simulations. $\mu$ increases from top to bottom. The dashed line is drawn in approximate correspondence to the mean field prediction for the pseudocritical $\mu$.

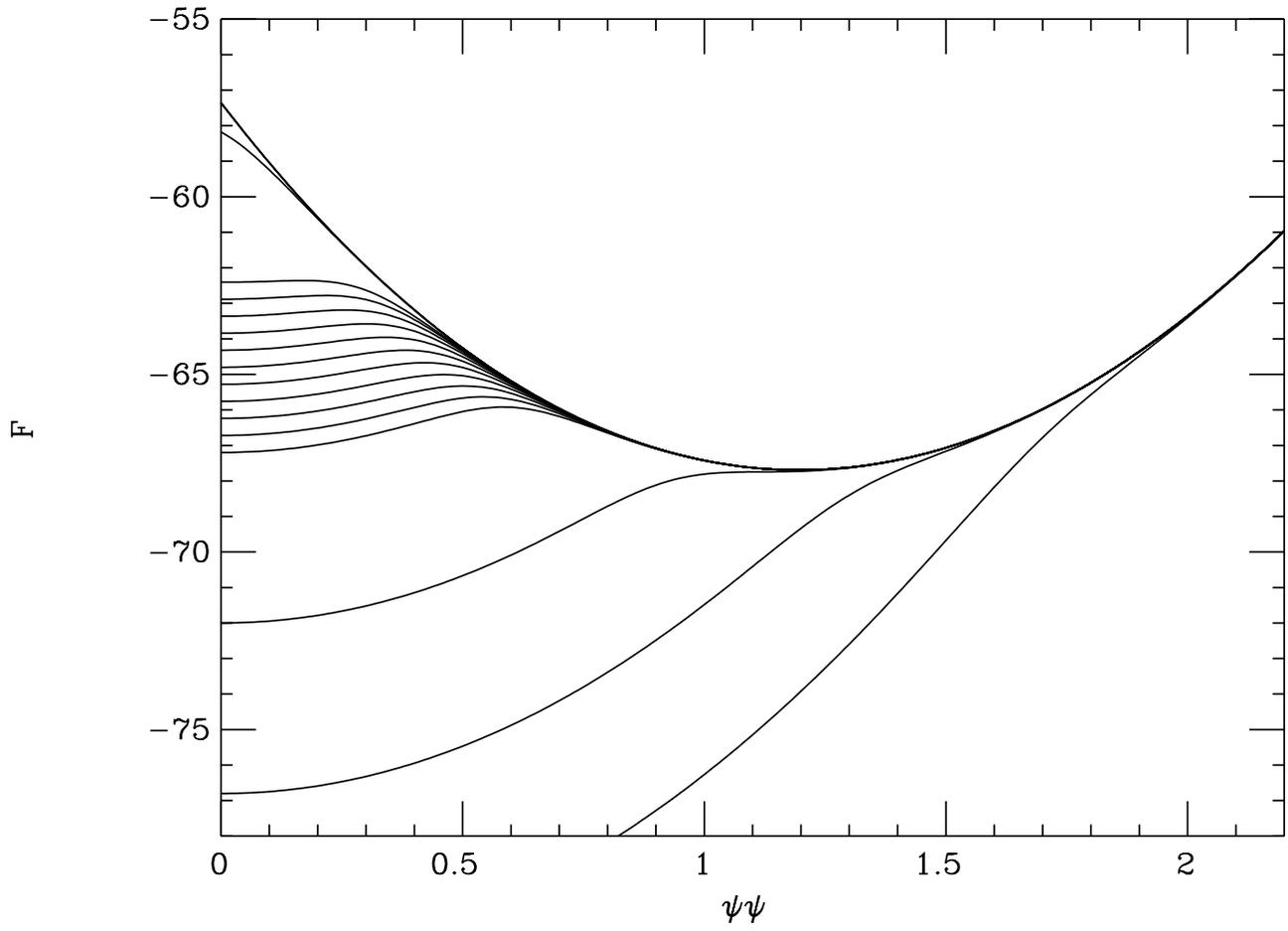

Figure 19b. As in Fig. 19a, but $m_q = 1.5$.

This, in turn, generates the secondary minimum around zero in the chiral condensate distribution, and produces the observed decline in the chiral condensate itself, which thus happens much before than the actual pseudocritical point $\mu^{*}$[1] To master the pathologies for $\mu > m_\pi/2$, and expose the physics of the chiral transition which is hidden deep inside the metastable region, we need to "enforce" the saddle point solution.

We thought that a natural possibility would be to increase the time extent of the lattice so the system could find the actual minimum : this is why we produced the 10 configurations on a $8^3 \times 32$ lattice mentioned at the beginning of this Section. However, the inversion time inside the metastable region remained very large, and the comparison between the results of the two lattices excluded strong volume effects.

This subject deserves further study. Comparisons with the behavior of other models can offer some guidance. In particular, we note that the zero temperature, finite density transition of the $U(1) \times U(1)$ Gross Neveu model, which has been successfully simulated [11], is second order [12] , i.e. does not have a metastable region. In SU(N) gauge models the transition turns first order for $N > 2$, while the SU(2) model has a second order transition [8]. First order transitions, with their "forbidden" metastable regions, and their numerical failures, seem associated with complex actions [2] , while models which can be successfully simulated have real actions, and undergo continuous transitions.

Clearly we need to uncover the physical reasons for the complex action at nonzero $\mu$, and/or to explore possible alternatives, such as the Hamiltonian formulation: we note that the authors of ref. [18] found that the critical chemical potential, computed at strong coupling in Hamiltonian lattice QCD, is indeed equal to the dynamical fermion mass, which, in the same scheme, is one third the nucleon mass [19].

---

[1] It is possible that the situation improves in the full model, since in that case the eigenvalue distribution does depend on the mass value, which can prevent the distribution itself from fluctuating randomly around the mass point on the real axes.

[2] This is consistent with the picture sketched above: only an eigenvalue distribution spread over the real axes (i.e. generated by a complex action) can produce the secondary minimum at $\bar{\psi}\psi = 0$, since this is due to the mass point falling inside the eigenvalue distribution.

## IV. CONCLUSIONS

We have performed an exhaustive study of quenched QCD at finite density in the scaling region, and in the strong coupling limit. We have measured the standard thermodynamic observables, the spectrum, and unphysical, ad-hoc observables meant to elucidate some of the peculiarities of the model. We confirmed the pathological behavior observed in the past for $\mu > m_\pi/2$. The onset at $m_\pi/2$ has been measured with great accuracy both at strong and weak coupling, and found to be independent of the lattice size, and the measurement technique. We think that our simulations in the scaling region have clearly exposed the pathologies at $m_\pi/2$. In particular, it has been shown that the early onset is indeed half the pion mass, which rules out the possibility of rescuing quenched, finite density QCD by using some refined nuclear matter models. However the accurate simulations performed in the scaling region do not suffice to clarify the very nature of the onset at $m_\pi/2$ – we have searched for hints of chiral symmetry restoration, both in the order parameter and in the spectrum, and found them to be very weak and inconclusive. The new simulations on large lattices have not greatly improved our understanding. We have also performed new, extensive simulations at strong coupling. We used different values of the bare quark mass, and a wide array of chemical potentials. We have also introduced new observables which helped shed more light on the pathologies. The analytic results available in that case offered a simple interpretation of the pathologies at $\mu = m_\pi/2$: the forbidden region of the simulations are to be associated with the metastable region in the effective potential. This suggests that the problems with finite density QCD could be solved if we could handle the metastable region.

Summarizing, the simulations of quenched QCD in the scaling region, large lattices, and at large $\mu$ value produce clear evidence of the major pathologies of the theory, but do not help to clarify the reasons behind these failures. The simulations in the strong coupling limit were particularly informative, especially when their results were analyzed in the context of available analytic treatments.

We wish to thank I. Barbour, G. Boyd, F. Karsch, B. Petersson, E. Mendel, G. Parisi and F. Niedermayer for conversations and correspondence.

This work was partially supported by NSF under grant NSF-PHY92-00148 and by DOE contract W-31-109-ENG-38. The simulations were done on the CRAY C–90's at PSC and NERSC.